\documentclass[a4paper,article]{cas-sc}
\usepackage{amsmath,amsfonts}
\usepackage{array}
\usepackage{makecell}
\usepackage[caption=false,font=normalsize,labelfont=sf,textfont=sf]{subfig}
\usepackage{textcomp}
\usepackage{stfloats}
\usepackage{url}
\usepackage{verbatim}
\usepackage{graphicx}
\usepackage{nomencl}
\usepackage{etoolbox}
\usepackage{ifthen}
\usepackage{booktabs}
\usepackage[ruled]{algorithm2e}
\usepackage{appendix}
\usepackage{multirow}
\usepackage{bm}
\usepackage{makecell}
\usepackage{algpseudocode}
\allowdisplaybreaks 
\usepackage[subtle,tracking=normal]{savetrees}
\usepackage{float}

\usepackage[numbers,sort&compress]{natbib}
\usepackage{nomencl}
\makenomenclature
\renewcommand{\nomgroup}[1]{%
    \ifthenelse{\equal{#1}{A}}{\item[\textbf{Indices and Sets}]}{%
    \ifthenelse{\equal{#1}{B}}{\item[\textbf{Abbreviations}]}{%
    \ifthenelse{\equal{#1}{C}}{\item[\textbf{Parameters}]}{%
    \ifthenelse{\equal{#1}{D}}{\item[\textbf{Decision Variables}]}{}}}}}
\usepackage{multicol}
\usepackage{framed} 
\usepackage{etoolbox} 
\usepackage{float}
\usepackage{placeins}
\usepackage{threeparttable} 

\def\tsc#1{\csdef{#1}{\textsc{\lowercase{#1}}\xspace}}
\tsc{WGM}
\tsc{QE}

\usepackage{amsthm}

\newtheorem{proposition}{Proposition}

\newtheorem{remark}{Remark}

\hypersetup{
    colorlinks=true,
    linkcolor=blue,
    citecolor=blue,
    urlcolor=blue
}

\begin{document}
\let\WriteBookmarks\relax
\def\floatpagepagefraction{1}
\def\textpagefraction{.001}
\let\printorcid\relax 

\shorttitle{\rmfamily K. Huang et al. Real-Time Peer-to-Peer Energy Trading}    

\shortauthors{\rmfamily K. Huang et al.}

\title[mode = title]{Real-Time Peer-to-Peer Energy Trading for Multi-Microgrids:\\ Improved Double Auction Mechanism and Prediction-Free\\Online Trading Approach}

\author[1]{Kaidi Huang}
\credit{Conceptualization, Methodology, Software, Validation, Visualization, Writing - original draft}

\author[1]{Lin Cheng}
\credit{Methodology, Supervision, Funding acquisition}

\author[2]{Yue Zhou}
\credit{Conceptualization, Methodology, Writing – review \& editing}


\author[1]{Fashun Shi}
\credit{Methodology, Resources, Supervision}

\author[1]{Yufei Xi}
\credit{Methodology, Writing – review \& editing}

\author[1]{Yingrui Zhuang}
\credit{Writing – review \& editing}

\author[3]{Ning Qi}
\ead{nq21767@columbia.edu} 
\credit{Methodology, Supervision, Writing – review \& editing}

\address[1]{Department of Electrical Engineering, Tsinghua University, Beijing 100084, China}
\address[2]{School of Electrical and Information Engineering, TianJin University, Tianjin, 300072, China}
\address[3]{Department of Earth and Environmental Engineering, Columbia University, New York, NY 10027, USA}

\cortext[1]{Corresponding author} 

\begin{abstract}
Peer-to-peer energy trading offers a promising solution for enhancing renewable energy utilization and economic benefits within interconnected microgrids. However, existing real-time P2P markets face two key challenges: high computational complexity in trading mechanisms, and suboptimal participant decision-making under diverse uncertainties. Existing prediction-based decision-making methods rely heavily on accurate forecasts, which are typically unavailable for microgrids, while prediction-free methods suffer from myopic behaviors. To address these challenges, this paper proposes an improved double auction mechanism combined with an adaptive step-size search algorithm to reduce computational burden, and a data-driven dual-reference online optimization (DDOO) framework to enhance participant decision-making. The improved mechanism simplifies bidding procedures, significantly reducing computational burden and ensuring rapid convergence to the market equilibrium. Additionally, the prediction-free DDOO framework mitigates myopic decision-making by introducing two informative reference signals. {\color{blue}Case studies on a 20-microgrid system demonstrate the effectiveness and scalability of the proposed mechanism and approach. The improved mechanism significantly decreases the computational time while increasing local energy self-sufficiency periods from 0.01\% to 29.86\%, reducing reverse power flow periods from 24.51\% to 3.96\%, and lowering average operating costs by 19.20\%.} Compared with conventional approaches such as Lyapunov optimization and model predictive control, the DDOO framework achieves a 10\%-13\% reduction in operating costs with an optimality gap of only 5.76\%.

\end{abstract}

\maketitle

\section{Introduction}

\subsection{Background and motivation}
A microgrid (MG) facilitates the integration and coordination of renewable energy sources (RES), energy storage (ES), distributed generation (DG), and diverse types of load. As RES penetration increases, MGs at the distribution level are evolving from passive consumers to active prosumers~\cite{zia2020microgrid}. In the traditional retail energy market, MGs purchase electricity from the main grid under a time-of-use (ToU) tariff and sell surplus energy back to the grid through a feed-in tariff (FiT) scheme~\cite{tushar2021peer}. However, low FiT revenues and restrictions on reverse power flows to the main grid substantially limit the profitability of prosumers and hinder the broader integration of RES.

Given this context, peer-to-peer (P2P) energy trading has emerged as a flexible market paradigm that enables interconnected MGs to directly exchange surplus energy locally. This enhances their economic benefits and promotes the local consumption of renewable energy sources (RES)~\cite{tushar2019grid}. However, existing real-time P2P market mechanisms face significant challenges, particularly the high computational complexity involved in frequent market clearing. Additionally, market participants are hindered by limited prediction-based decision-making methods due to the difficulty in obtaining accurate forecasts, while prediction-free approaches often exhibit myopic behaviors. Therefore, developing efficient market mechanisms with reduced computational complexity and prediction-free decision-making frameworks is essential to mitigate myopic behaviors in real-time P2P energy trading.

\subsection{Literature review}
(1) P2P market structures and pricing mechanisms

\newpage
\nomenclature[A]{$r/N^r$}{Indices for MGs$/$total number of MGs}
\nomenclature[A]{$i/b$}{Indices for buses$/$branches}
\nomenclature[A]{$s/t$}{Indices for historical scenarios$/$time period}
\nomenclature[A]{$\bm{\Omega}_{{B}}/\bm{\Omega}_{{T}}$}{Sets for buses$/$time period}
\nomenclature[A]{$Br(i)$}{Set of the branches that connect to bus $i$}
\nomenclature[A]{$Br(i,j)$}{Set of the branch between bus $i$ and $j$}

\nomenclature[B]{ASSA}{Adaptive step-size search algorithm}
\nomenclature[B]{ES$/$GES}{$\text{Energy storage}/$generalized energy storage}
\nomenclature[B]{MG$/$MPC}{$\text{Microgrid}/$model predictive control}
\nomenclature[B]{RES$/$SoC}{$\text{Renewable energy sources}/$state of charge}
\nomenclature[B]{DDOO}{Data-driven dual-reference online optimization}
\nomenclature[B]{DG$/$VES}{$\text{Distributed generator}/$virtual energy storage}
\nomenclature[B]{ToU$/$FiT}{$\text{Time-of-use}/$feed-in tariff}
\nomenclature[B]{So$/$RO}{Stochastic optimization$/$robust optimization}
\nomenclature[B]{DRO}{Distributionally robust optimization}
\nomenclature[B]{P2P$/$P2G}{$\text{Peer-to-peer}/$peer-to-grid}
\nomenclature[B]{ADMM}{Alternating direction method of multipliers}
\nomenclature[B]{IRT$/$RPB}{$\text{Informative reference trajectory}/$real-time price benchmark}
\nomenclature[C]{$\Delta t/a$}{Unit dispatch interval$/$cost coefficient of DG}
\nomenclature[C]{$\lambda_t^{\text{ToU}}/\lambda_t^{\text{FiT}}$}{$\text{ToU price}/$FiT at time $t$}
\nomenclature[C]{$\overline{P}_{t}^\mathrm{G,c}$}{Maximal charging power of GES}
\nomenclature[C]{$\overline{P}_{t}^\mathrm{G,d}$}{Maximal discharging power of GES}
\nomenclature[C]{$R_b/X_b$}{$\text{Resistance}/$reactance of branch $b$}
\nomenclature[C]{$\overline{V}^{\mathrm{BUS}}$}{Maximal voltage deviation}
\nomenclature[C]{${V}^{\mathrm{B}}$}{Voltage of the substation bus}
\nomenclature[C]{$\underline{SoC}_{t}/\overline{SoC}_{t}$}{$\text{Lower}/$upper bounds of GES SoC}
\nomenclature[C]{$\eta^{\text{c}}/\eta^{\text{d}}$}{$\text{Charging}/$discharging efficiency of GES}
\nomenclature[C]{$c_{\mathrm{d}}^\mathrm{G}/c_{\mathrm{c}}^\mathrm{G}$}{Cost coefficients of GES}
\nomenclature[C]{$\mathcal{E}/E$}{Self-discharge rate$/$capacity of GES}
\nomenclature[C]{$\pi_{t}$}{Baseline consumption of GES}
\nomenclature[C]{$\underline{P}^{\mathrm{DG}}/\overline{P}^{\mathrm{DG}}$}{$\text{Lower}/$upper power bounds of DG}
\nomenclature[C]{$\mu/\gamma$}{$\text{Mean}/$standard deviation of the distribution}
\nomenclature[C]{$\varepsilon$}{Confidence level of chance constraint}
\nomenclature[D]{$P_{t,r}^{\rm EX}$}{Bidding quantity of MG $r$}
\nomenclature[D]{$P_{t}^{\rm grid}$}{Power exchange with utility grid}
\nomenclature[D]{$P_{t}^{\text{G,c}}$/$P_{t}^{\text{G,d}}$}{$\text{Charging}/$discharging power of GES}
\nomenclature[D]{$P_{b,0,t}^{\mathrm{PF}}/Q_{b,0,t}^{\mathrm{PF}}$}{$\text{Active}/$reactive power that flows on the lateral branch of branch $b$}
\nomenclature[D]{$\lambda_t^{\text{P2P}}$}{P2P market trading price}
\nomenclature[D]{$P_{b,t}^{\mathrm{PF}}/Q_{b,t}^{\mathrm{PF}}$}{$\text{Active}/$reactive power that flows on branch $b$}
\nomenclature[D]{$V_{i,t}^{\mathrm{BUS}}$}{Voltage at bus $i$}
\nomenclature[D]{$P_{t}^{\text{DG}}/P_{t}^{\text{R}}/P_{t}^{\text{L}}/\ell_{t}$}{$\text{DG power}/\text{RES power}/\text{load}/\text{net load}$}
\begin{framed}
    \begin{multicols}{2}
        \printnomenclature
    \end{multicols}
\end{framed}

 According to the organization of trading and information exchange, the P2P market structure in the existing works is typically categorized into three types: centralized market, decentralized market, and community-based market~\cite{tushar2021peer}. Centralized P2P markets adopt a hierarchical structure, where a central coordinator manages the entire trading process. Specifically, the central coordinator collects operational parameters from all market participants and optimizes energy schedules to maximize overall social welfare~\cite{zhong2020cooperative,chen2020peer,alizadeh2024cooperative}. Although centralized structures are straightforward to implement, they have several drawbacks, including high computational complexity with increasing participants, privacy issues due to full information disclosure, and challenges in ensuring fair pricing and equitable benefit allocation. To overcome these limitations, extensive research attention has shifted toward the decentralized P2P market structure, where market participants directly trade energy with each other without relying on a central coordinator~\cite{mehdinejad2022peer}. This decentralized architecture significantly enhances participant privacy by granting peers full autonomy over their own trading decisions. In practice, various distributed optimization algorithms, such as the alternating direction method of multipliers (ADMM)~\cite{wei2021optimal,shi2023distributed,wu2023multi,li2021data,kim2019direct,zheng2025real,liu2023online,hou2024distributed}, dual gradient-based methods~\cite{feng2022peer,khorasany2019decentralized}, and consensus-based approaches~\cite{sorin2018consensus}, have been extensively adopted to facilitate effective peer-to-peer transactions. Among these, the Nash bargaining model~\cite{wu2023multi,wei2021optimal,kim2019direct} is widely used as a pricing mechanism, in which the P2P market-clearing problem is formulated as a social welfare maximization model. This problem is subsequently decomposed into smaller subproblems and iteratively solved using distributed optimization techniques, most notably ADMM. Despite its advantages, the decentralized P2P market faces several critical challenges, including the inability of distributed iterative methods to guarantee global optimality~\cite{tushar2021peer,wu2023hierarchical,zheng2024multi}, convergence and computational efficiency issues as the number of prosumers grows~\cite{tushar2021peer}, and increased complexity and communication overhead due to extensive infrastructure requirements~\cite{zhu2022peer,feng2024communication}.

In contrast, community-based P2P markets represent a hybrid structure positioned between centralized and fully decentralized markets, where a third-party coordinator facilitates trading by managing limited information exchange (e.g., collecting prosumers’ bid/ask prices and quantities, issuing market-clearing prices), rather than directly controlling transactions~\cite{zheng2024multi,zhu2022peer,he2020community,qiu2021scalable}. Compared to the decentralized market, it enhances efficiency by reducing the complexity of fully distributed iterative computation while maintaining prosumer autonomy. This structure addresses scalability and privacy concerns in the centralized market, as well as optimality, computational efficiency, and communication overhead challenges in the decentralized market~\cite{zheng2024multi,zhu2022peer}. One of the core issues of the community market is employing an attractive pricing mechanism. Previous studies have explored two primary pricing approaches: uniform pricing and double auction schemes. In uniform pricing mechanisms, such as the supply-demand ratio~\cite{liu2017energy}, mid-market rate~\cite{wu2023hierarchical,qiu2021scalable}, and bill sharing method~\cite{long2017peer}, prices are centrally determined based on the participants’ net demand and selling/buying prices of the grid. While these schemes are easy to implement, they limit participants' ability to quote prices and the amount of energy autonomously, and the resulting energy trading and control decisions may not be utility-maximizing~\cite{zheng2024multi}. In contrast, double auction ~\cite{zheng2024multi,guerrero2018decentralized,cui2024consortium} allows participants to submit both price and quantity bids/asks, enabling individualized pricing for different transactions. This approach fosters market competition and enhances potential profits. By allowing buyers and sellers to compete simultaneously, the double auction helps align prices with actual market supply and demand, thereby facilitating more efficient resource allocation. However, the primary challenge lies in the computationally intensive and time-consuming high-dimensional bidding process, which can significantly impede trading efficiency~\cite{wu2023hierarchical,liu2025reinforcement}.

{\color{blue}
In traditional double auction mechanisms, participants are required to submit high-dimensional bids~\cite{zheng2023energy,qi2025privacy} in each trading period, i.e., submitting multiple price–quantity pairs to accurately reflect the amount of energy they are willing to trade at different market prices. This imposes a heavy computational burden on prosumers. To address this challenge, some studies have proposed \textit{iterative double auction} mechanisms, where each participant initially submits only a single price–quantity bid and then gradually adjusts this bid across multiple iterations until the market converges to equilibrium~\cite{zhang2021peer,xu2021iterative, hou2024novel}. This simplification substantially reduces the bidding complexity for participants. However, for participants, especially when time-coupling entities such as ES are involved, designing appropriate initial bids and effective adjustment rules under complex uncertainties remains difficult, and the existing literature often relies on simplified or unrealistic assumptions. In \cite{zhang2021peer}, each prosumer is assumed to have perfectly accurate one-hour-ahead forecasts, while ES is not considered in the model. In \cite{xu2021iterative}, the arbitrage capability of ES is ignored. The charging and discharging decisions are not explicitly optimized in the bidding stage; instead, ES is mainly used after market clearing to absorb unsold surplus or supply unmet deficits. In \cite{hou2024novel}, ES is incorporated into bidding by defining a quantity space that depends on the state-of-charge (SoC) together with predicted PV generation and base loads. Each prosumer then solves a mixed-integer nonlinear program to decide its role, bidding quantity, and price, while the auctioneer solves a mixed-integer linear program for winner determination. This design raises severe scalability concerns, since both prosumers and the auctioneer face heavy computational burdens. Moreover, its iterative rule is simplistic—unmatched bidders adjust prices by a fixed step—and no theoretical convergence guarantee is provided. It should be noted that \cite{zhang2021peer,xu2021iterative,hou2024novel} utilize predicted generation and load values as inputs, but do not explicitly account for prediction errors or system uncertainties.
}

\begin{table*}[!ht]
\footnotesize\rmfamily
  \centering
  \begin{threeparttable}
  \caption{\rmfamily Comparison of this paper with recent P2P energy trading literature.}
  \setlength{\tabcolsep}{0.7mm}{
      \begin{tabular}{c c c c c c c c}
    \toprule
    Ref.  & \makecell{Market Architecture\\/ Pricing Mechanism} & \makecell{Distributed\\Algorithm}& \makecell{Time-coupling\\entities} & \makecell{System\\Uncertainty} & Prediction-Free &\makecell{Real-time\\Operation} & \makecell{Simulation System\\/ Time Resolution}\\
    \midrule
    \cite{zhong2020cooperative}  & Cen. / Nash Bargaining & — & — & — & — & — & 15 prosumers / 1-h\\
    \cite{chen2020peer}  & Cen. / Stochastic game & — & — &  $\checkmark$ & — & — & 5 prosumers / 1-h\\
    \cite{alizadeh2024cooperative}  & Cen. / Nash Bargaining & — & $\checkmark$& $\checkmark$ & — & — & 9 prosumers / 1-h\\
    \cite{feng2022peer}  & Dec. / Nash Bargaining & LR-M + GF-DA& — &  — & — & — & 52 prosumers / 10-min\\
    \cite{wei2021optimal}  & Dec. / Nash Bargaining & ADMM & $\checkmark$ & $\checkmark$& — & — & 4 MGs / 1-hour\\
    \cite{shi2023distributed}  & \makecell{Dec. / MC-based pricing} & ADMM+MPC  & $\checkmark$ &  $\checkmark$ & — & $\checkmark$& 5 prosumers / 15-min\\
    \cite{wu2023multi}  & Dec. / Nash Bargaining & bi-level nested ADMM &  $\checkmark$ &  $\checkmark$ & — & — & 4 MGs / 1-h\\
    \cite{lou2025privacy}  & Dec. / Nash Bargaining & \makecell{PAC} &  $\checkmark$ &  $\checkmark$ & — & — & 10 agents / 1-h\\
    \cite{li2021data}  & \makecell{Dec. / — (Implicit pricing\\via ADMM)} & ADMM  & $\checkmark$ &  $\checkmark$ & — &  — & 4MGs / 1-h\\
    \cite{zheng2025real}  & \makecell{Dec. / — (Implicit pricing\\via ADMM)} &  Lyapunov+ADMM     & $\checkmark$ &$\checkmark$ & $\checkmark$ (Myopic) & $\checkmark$  & 3 energy hubs / 1-h\\
    \cite{liu2023online}  & \makecell{Dec. / — (Implicit pricing\\via ADMM)} &  Lyapunov+ADMM  & $\checkmark$ & $\checkmark$ & $\checkmark$ (Myopic) & $\checkmark$  & 14 prosumers / 15-min\\
    \cite{zhao2023stackelberg}  & Comm. / Stackelberg game & Multi-agent DRL  & $\checkmark$ &  $\checkmark$ & — & $\checkmark$ & 4MGs / 1-h\\
    \cite{wu2023hierarchical}  & Comm. / Mid-market rate & Multi-agent DRL  & $\checkmark$ & — & — & — & 8 MGs / 4-hour\\
   \cite{zheng2024multi}  & Comm. / Double auction & Multi-agent RL  & $\checkmark$ &   $\checkmark$ & —& $\checkmark$ & \makecell{10 prosumers / 1-h}\\
    \cite{zhu2022peer}  & Comm. / Double auction &  Lyapunov-based control  & $\checkmark$ & $\checkmark$ & $\checkmark$ (Myopic) & $\checkmark$  & 10 prosumers / 1-h\\
    \cite{zhang2021peer}  & \textcolor{blue}{Comm. / Iterative double auction} &  \textcolor{blue}{Heuristic iteration}  & \textcolor{blue}{—} & \textcolor{blue}{—} & \textcolor{blue}{—} & \textcolor{blue}{$\checkmark$}  & \textcolor{blue}{30 prosumers / 1-h}\\
    \cite{xu2021iterative}  & \textcolor{blue}{Comm. / Iterative double auction} &  \textcolor{blue}{AMSA}  & \textcolor{blue}{$\checkmark$} & \textcolor{blue}{—} & \textcolor{blue}{—} & \textcolor{blue}{$\checkmark$}  & \textcolor{blue}{5 prosumers / 1-h}\\
    \cite{hou2024novel}  & \textcolor{blue}{Comm. / Iterative double auction} &  \textcolor{blue}{MINLP-based iteration}  & \textcolor{blue}{$\checkmark$} & \textcolor{blue}{—} & \textcolor{blue}{—} & \textcolor{blue}{$\checkmark$}  & \textcolor{blue}{30 prosumers / 5-s}\\
    \makecell{This\\paper}   & Comm. / Improved double auction & \makecell{ASSA+DDOO} & $\checkmark$ &$\checkmark$ & $\checkmark$ & $\checkmark$& 20 MGs / 5-min\\
    
    \bottomrule
    \end{tabular}%
    }\label{literature review}
    \begin{tablenotes}
\item $\checkmark$: The item is considered. —: The item is not considered; Cen. : Centralized market; Dec. : Decentralized market; Comm. : Community market. LR-M: Lagrangian relaxation-based method; GF-DA: Generalized fast dual ascent; MC-based pricing : Marginal cost-based pricing; PAC: proximal atomic coordination; \textcolor{blue}{AMSA: Auction market self-adaption algorithm; MINLP: Mixed-integer nonlinear program.}
\end{tablenotes}
\end{threeparttable}\vspace{-0.5cm}
\end{table*}%

(2) Decision-making under uncertainty in P2P energy trading

With the high penetration of RES into power systems, decision-making under uncertainty has become a critical challenge for participants in P2P energy trading. Existing research can be first categorized based on whether time-coupling entities, primarily ES, are considered: single-period and multi-period P2P energy trading. Single-period P2P energy trading ignores time-coupling entities and primarily focuses on decentralized trading mechanisms~\cite{paudel2020peer}, incorporating physical network constraints~\cite{zhong2020cooperative}, and accelerating distributed trading algorithms~\cite{feng2022peer}. Although these approaches provide valuable insights, they cannot be directly extended to multi-period scenarios. Specifically, when time-coupling entities are present, independently optimizing decisions at each time slot—akin to a greedy strategy—often results in suboptimal performance over the full operational horizon~\cite{liu2023online}. 

In contrast, multi-period P2P trading explicitly considers time-coupling entities in multi-period P2P trading problems, further categorized based on how uncertainties are addressed: deterministic, prediction-based, and prediction-free online optimization models. Deterministic optimization models, adopted in~\cite{bo2023peer,wu2023hierarchical,cao2022efficient,javadi2022transactive}, analyze market equilibrium and economic benefits with exact problem data, neglecting system uncertainties. However, uncertainties in renewable generation, load profiles, and energy prices significantly impact the economic and reliable operation of MGs and cannot be overlooked in practice. Extensive research employs prediction-based approaches to handle uncertainty, including stochastic optimization (SO)~\cite{alizadeh2024cooperative,wang2023distributed}, robust optimization (RO)~\cite{khodoomi2023robust}, and distributionally robust optimization (DRO)~\cite{zhang2023distributionally}, chance-constrained optimization~\cite{qi2023chance}. These methods primarily address uncertainties at the day-ahead planning stage but often fail to adapt effectively to real-time environmental changes. This limitation motivates the exploration of multi-period rolling horizon dispatch methods that rely on continuously updated forecasts using advanced techniques such as model predictive control (MPC)~\cite{shi2023distributed}, dynamic programming~\cite{li2021multi}, and reinforcement learning~\cite{zhao2024automatic}. However, their performance largely depends on forecast accuracy. In practice, precise predictions of RES outputs and market prices are typically unavailable or unreliable, especially given the challenges of high-quality RES prediction in small regions~\cite{zheng2025real} and the inherent volatility and unpredictability of market prices. Consequently, existing prediction-based approaches may result in poor economic performance or face feasibility issues in some cases~\cite{wang2023online}.

To address these challenges, recent studies have explored online algorithms that reduce reliance on forecasting models. Among them, Lyapunov optimization has gained significant attention as a prediction-free technique originally developed for stochastic optimization in queuing and communication networks~\cite{neely2010stochastic}. By introducing a Lyapunov function, it reformulates the long-term time-average cost minimization into a sequential drift-plus-penalty problem, eliminating the need for future predictions. Due to the conceptual analogy between queue accumulation and ES dynamics, Lyapunov optimization has been widely applied to real-time operational problems involving ES~\cite{zheng2025real,shi2015real,liu2023online}. Although Lyapunov optimization ensures long-term operational stability of ES by using the drift term to keep the storage SoC bounded and stable, and employs the penalty term to minimize immediate operating costs, it inherently exhibits a myopic nature. Due to the inter-temporal coupling constraints of ES, decisions based solely on instantaneous system states without any form of long-term guidance~\cite{qi2025long} can result in locally optimal but globally suboptimal storage dispatch. This short-sightedness arises because the algorithm lacks either precise predictions or an informative long-term guidance to anticipate future operating conditions.

\subsection{Research gap}

Existing literature is summarized in Table~\ref{literature review}. Despite the extensive literature, several critical research gaps remain insufficiently addressed.

(1) Conventional double auction mechanisms face significant challenges, as participants typically submit high-dimensional bids. This bidding approach has been widely adopted in practical electricity markets such as Pennsylvania-New Jersey Maryland (PJM)~\cite{PJM2022manual}, California independent system operator (CAISO)~\cite{CAISO2023RTM}, and Australian energy market operator (AEMO)~\cite{AEMO2022ESS}, due to its ability to capture market dynamics, promoting competition and enhance participant profitability~\cite{liu2025reinforcement}. \textbf{However, in real-time P2P markets characterized by high trading frequency and short clearing intervals (e.g., five minutes interval), such complex bidding processes significantly increase computational complexity and impose a substantial computational burden on market participants.} \textcolor{blue}{Although iterative double auction mechanisms have been proposed to mitigate this issue by simplifying bids to a single price–quantity pair updated across iterations~\cite{zhang2021peer,xu2021iterative,hou2024novel}, these methods still rely on simplified assumptions (e.g., perfect forecasts or neglect of storage arbitrage), lack explicit consideration of uncertainty, or face scalability limitations.} {\color{blue}\textbf{Moreover, conventional double auctions aggregate discrete price–quantity bids, which often fail to yield an exact market equilibrium.}} The resulting supply–demand imbalance must typically be absorbed by the external grid, reducing the overall benefits of P2P trading. {\color{blue}In this paper, we address this gap by developing an improved double auction mechanism that eliminates high-dimensional price bidding, achieves exact supply--demand balance, and sharply reduces participant-side computations, thereby ensuring both market efficiency and scalability in real-time P2P settings.}

(2) Existing approaches to handling uncertainty in P2P energy trading exhibit notable limitations. \textbf{Although prediction-based methods, including SO, RO, DRO, and MPC, have been widely studied, their effectiveness heavily depends on accurate forecasts, which are often unavailable or unreliable in small-scale systems.} Meanwhile, Lyapunov optimization, a prediction-free online method, avoids the need for precise forecasts; \textbf{however, due to its inherent myopic nature and lack of long-term guidance, it risks producing ES dispatch decisions that are locally optimal yet significantly deviate from optimality over a longer horizon.} {\color{blue}To overcome these limitations, this paper introduces a prediction-free data-driven dual-reference online optimization (DDOO) framework. By incorporating two reference signals—an informative reference trajectory that embeds long-term scheduling experience and a real-time price benchmark that reflects opportunity values of remaining capacity—our approach effectively mitigates the short-sightedness of storage scheduling and achieves performance close to the hindsight optimum.}

\subsection{Contributions}

{\color{blue}
This paper advances real-time P2P trading by (i) designing an \emph{improved double auction} that \emph{achieves exact supply--demand balance while greatly reducing participant-side computations}, and (ii) proposing a \emph{prediction-free} online decision framework that \emph{mitigates myopic storage operations}. Together, the mechanism and the optimization framework deliver practical feasibility and stronger economic performance at a 5-minute resolution.

\textbf{(1) Market mechanism --- improved double auction with exact clearing and low computational burden.}
We design an improved double auction mechanism in which the P2P operator iteratively updates the P2P price and participants submit only quantity bids. 
Unlike conventional double auctions (involving high-dimensional bidding that requires participants to submit multiple price--quantity pairs) that often yield only an approximate balance due to discrete bids and impose heavy computational effort on participants, the proposed mechanism \emph{converges to an exact clearing price} that balances supply and demand \emph{while sharply reducing participant-side computations}. In our 60-day simulation, the mechanism eliminates the need for high-dimensional price bidding and therefore achieves a 4.84–9.68-fold speedup per interval compared with conventional double auction mechanisms.

\textbf{(2) Solution algorithm --- ASSA with finite-step convergence guarantees.}
In the proposed market design, the P2P operator employs an adaptive step-size search algorithm (ASSA) to iteratively update the trading price until supply and demand are exactly balanced. We prove the \emph{existence and uniqueness} of the clearing price and \emph{finite-step convergence} to it, independent of the initialization. Empirically, ASSA enables the mechanism to converge in only 2.07 iterations on average per 5-minute interval over 60 days, thereby supporting real-time deployment.

\textbf{(3) Decision-making framework --- prediction-free DDOO mitigating myopic decisions.}
We propose a prediction-free, DDOO that incorporates two references:  
(i) an \emph{informative reference trajectory} (IRT) for SoC to provide long-term guidance, and  
(ii) a \emph{real-time price benchmark} (RPB) to embed the opportunity value of remaining storage capacity. This directly tackles the short-sightedness of storage dispatch without requiring load/price forecasts. Across 60 days, the framework improves operating economics (e.g., average MG cost reduction by 19.2\%), increases self-sufficiency periods (0.01\% $\rightarrow$ 29.86\%), and reduces reverse power-flow periods (24.51\% $\rightarrow$ 3.96\%), while maintaining a small optimality gap (5.76\%) relative to the hindsight optimum.}

\subsection{Paper Organization}

We organize the remainder of the paper as follows. Section~\ref{mechanism} introduces the market mechanism for real-time P2P energy trading. Section~\ref{game} models the market behavior using a non-cooperative Nash game, proposes the prediction-free DDOO framework for individual microgrids, and provides theoretical analysis for market equilibrium and the ASSA solution method. Section~\ref{case} presents comprehensive numerical case studies to verify the effectiveness of the proposed approach. Finally, Section~\ref{conclusion} concludes the paper and discusses future work directions.

\section{Market mechanism for real-time P2P energy trading}~\label{mechanism}

\subsection{Market architecture}

\begin{figure}[htbp]
  \footnotesize\rmfamily   \setlength{\abovecaptionskip}{-0.1cm}  
    \setlength{\belowcaptionskip}{-0.1cm} 
  \begin{center}  \includegraphics[width=0.5\columnwidth]{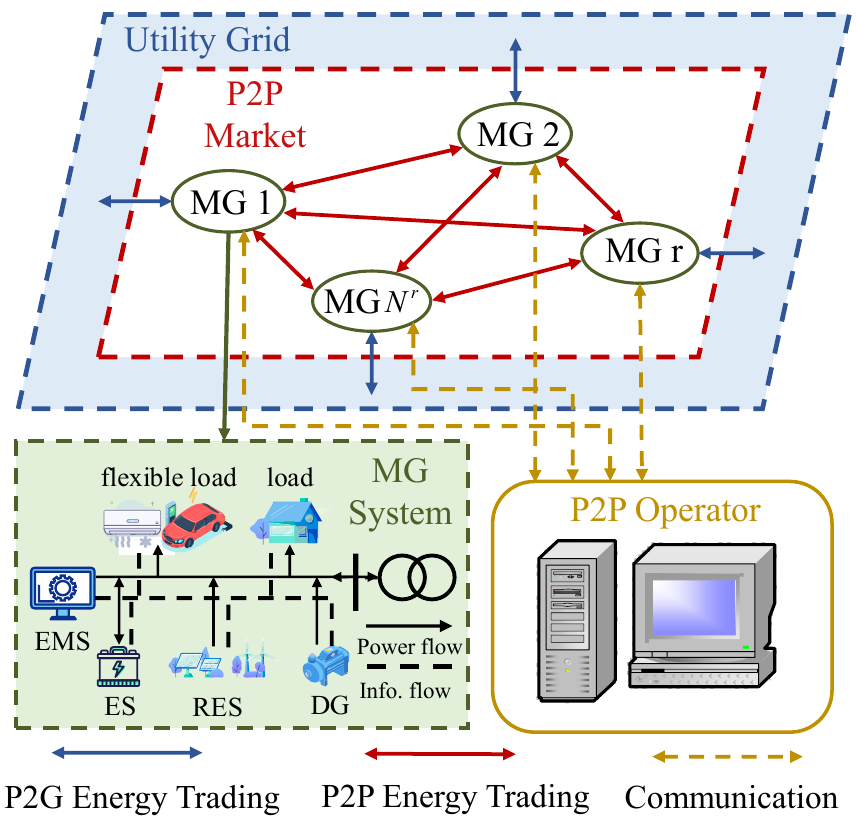}
     \caption{\rmfamily Market architecture of P2P energy trading.}\label{architecture}
  \end{center}
  \vspace{-2.5em}
\end{figure}
Fig.~\ref{architecture} illustrates a typical community market architecture, consisting of multiple interconnected MGs integrated with the utility grid. These MGs are allowed to trade energy fairly via a centralized sharing platform managed by a non-profit P2P market operator. Each MG functions as a prosumer, capable of buying and selling energy either through P2P transactions or peer-to-grid (P2G) interactions. Specifically, P2G transactions follow conventional tariffs, employing ToU tariff for energy purchases from the utility grid and FiT for energy sold back to the grid. In contrast, the P2P market transactions are managed by the P2P operator who centrally collects bid/ask information (price and quantity) from participating MGs and subsequently disseminates clearing prices, without requiring detailed information about each MG's internal cost functions. Consequently, the computational overhead and associated costs remain low, rendering this architecture suitable for large-scale markets.

Regarding the market-clearing price, two essential points should be highlighted: (1) The market-clearing price in the P2P market must be uniform for buyers and sellers at each clearing interval. According to~\cite{anoh2019energy},\cite{yan2020distribution}, each producer’s selling price will be the same and equal to the purchasing price of each consumer when the perfectly competitive P2P market is in equilibrium. (2) The clearing price must be bounded between the current ToU and FiT tariffs, thereby incentivizing MGs to prioritize P2P transactions over P2G interactions. Specifically, selling electricity within the P2P market at the clearing price ensures higher revenue than selling at FiT, and purchasing from peers at the market-clearing price is more economical than purchasing from the utility grid under ToU tariffs. Only when the local community trading cannot balance supply and demand does the MG trade with the utility grid. Accordingly, the P2P market promotes local energy balancing at the distribution level.

\subsection{Improved double auction mechanism}
\begin{figure}[htbp]
  \footnotesize\rmfamily   \setlength{\abovecaptionskip}{-0.1cm}  
    \setlength{\belowcaptionskip}{-0.1cm} 
  \begin{center}  \includegraphics[width=0.5\columnwidth]{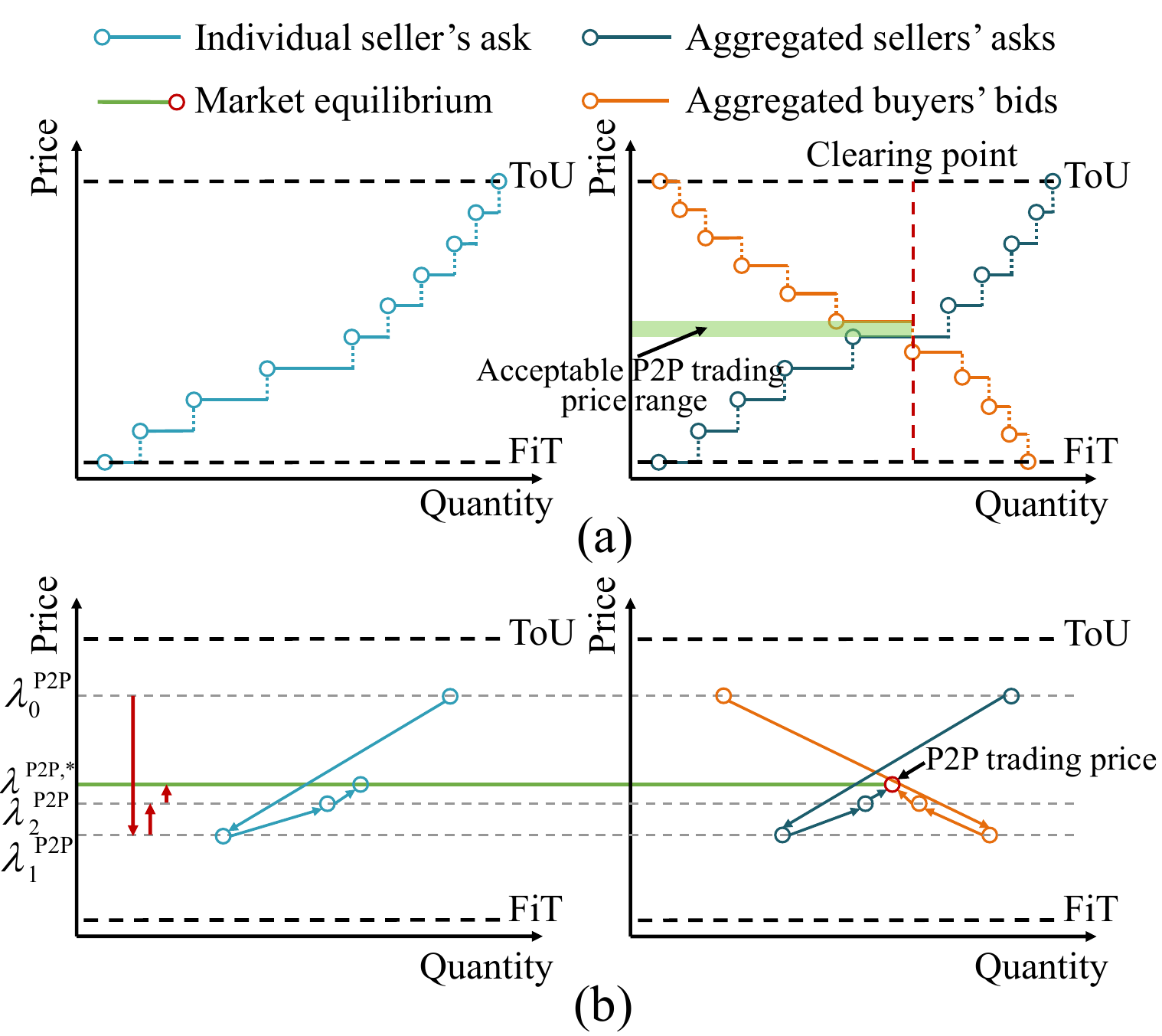}
     \caption{\rmfamily Comparison between (a) the traditional double auction mechanism and (b) the proposed mechanism.}\label{scheme}
  \end{center}
  \vspace{-2.5em}
\end{figure}

The double auction mechanism allows multiple buyers and sellers to independently determine their bids and asks based on predefined decision-making rules, without prior knowledge of other participants' strategies. The traditional double auction process is illustrated in Fig.~\ref{scheme} (a).

\textit{Step1.} Each participant generates bidding information based on their own optimization strategies and submits it to the P2P operator. Specifically, the bidding information consists of multiple price-power pairs $(\hat{\lambda},\hat{p})$, as shown in the left subfigure of Fig.~\ref{scheme} (a).

\textit{Step2.} The P2P operator aggregates all sellers' asks and buyers' bids to generate the aggregated supply and demand curve. The aggregated asks are sorted in ascending order, while the aggregated bids are sorted in descending order. The intersection of these two curves determines the market equilibrium point, as illustrated in the right subfigure of Fig.~\ref{scheme} (a). The green-shaded area represents the acceptable P2P trading price range, and various mechanisms exist in the literature to determine the final clearing price~\cite{tushar2019grid}.

However, traditional double auction mechanisms exhibit two key shortcomings: (1) The high-dimensional bidding process imposes significant computational complexity. Specifically, if the number of price-power pairs is denoted as 
$D$, each participant must execute their optimization strategy $D$ times every five minutes, substantially increasing computational requirements. (2) Since participants' bids and asks consist of discrete price-power pairs, the obtained clearing point does not represent a perfect market equilibrium, meaning that supply and demand are not exactly balanced. Consequently, an additional allocation process is required. A smaller bidding dimension $D$ results in a greater deviation of the clearing point from the true market equilibrium.

The proposed mechanism effectively addresses these challenges. The core principle of the double auction mechanism is to determine a clearing price at which the total quantity supplied by sellers matches the total quantity demanded by buyers, achieving supply-demand balance. To this end, we modify the auction procedure, as illustrated in Fig.~\ref{scheme} (b).

\textit{Step 1.} The P2P operator announces a trading price (initialized as $\lambda_0^{\text{P2P}}$) to all participants. Each seller and buyer then submits their bid/ask quantity at this price to the P2P operator.

\textit{Step 2.} The P2P operator adjusts the trading price based on the supply-demand imbalance and redistributes the updated price to all participants. This process repeats until supply and demand are balanced.

Compared to the traditional mechanism, the proposed approach ensures that the final clearing price is a deterministic value rather than a range, representing the exact market equilibrium point. However, the feasibility of this mechanism depends on its ability to efficiently converge to a market-clearing price. To this end, Section~\ref{game} introduces ASSA, which enables rapid convergence to the equilibrium price. Assuming the mechanism achieves balance after $iter$ price adjustments, the resulting computational speedup of the improved mechanism is given by:

\begin{equation}
\text { Speedup Factor }=\frac{D}{iter+1}\label{speed_up}
\end{equation}
 where $D$ represents the number of price-power pairs in the traditional double auction mechanism.
\begin{remark}[P2P and P2G Relationship]
In a P2P market, supply and demand are not always perfectly balanced. When the clearing price \(\lambda_t^{\text{P2P}}\) falls within the range \(\lambda_t^{\text{FiT}} < \lambda_t^{\text{P2P}} < \lambda_t^{\text{ToU}}\), the P2P market achieves internal supply-demand equilibrium, eliminating the need for peer-to-grid (P2G) energy trading with the utility grid. However, if the clearing price reaches \(\lambda_t^{\text{FiT}}\), it indicates that supply exceeds demand within the P2P market, requiring excess energy to be sold to the utility grid. Conversely, if the clearing price equals \(\lambda_t^{\text{ToU}}\), it signifies that demand surpasses supply, necessitating energy purchases from the utility grid. This relationship is mathematically expressed as follows:

\begin{equation}
\begin{aligned}
&\text{if } \lambda_t^{\text{FiT}} < \lambda_t^{\text{P2P}} < \lambda_t^{\text{ToU}},  \quad P_t^{\text{grid}} = 0. \\
&\text{if } \lambda_t^{\text{P2P}} = \lambda_t^{\text{ToU}}, \quad P_t^{\text{grid}} \leq 0. \\
&\text{if } \lambda_t^{\text{P2P}} = \lambda_t^{\text{FiT}}, \quad P_t^{\text{grid}} \geq 0.
\end{aligned}\label{condition}
\end{equation}

\end{remark}

\section{Real-time P2P energy trading}~\label{game}
To analyze the real-time P2P trading process and determine the market equilibrium, it is essential to model the trading behavior of each MG. Without loss of generality, this study focuses on a perfectly competitive P2P market, where each participant aims to maximize individual profit, and MGs are free to make self-interested decisions. Accordingly, game theory provides a suitable framework for modeling MG behavior and analyzing market equilibrium.

In this section, the bidding behavior of MGs is modeled using a non-cooperative Nash game, and the corresponding mathematical formulation of market equilibrium is derived. A theoretical proof of the existence and uniqueness of the equilibrium is provided. To enhance MG profitability in market operation, a prediction-free two-stage coordinated optimization framework is proposed, which effectively addresses the myopic nature of online optimization algorithms. Finally, an ASSA is developed, enabling the P2P operator to efficiently identify the market equilibrium, thus making it well-suited for real-time P2P markets with five-minute clearing intervals.

\subsection{Non-cooperative Nash game formulation}\label{game_formulation}
In the real-time P2P market, trading occurs every five minutes. At each time step $t$, the trading process can be modeled as a non-cooperative Nash game. Specifically, the Nash game $\mathcal{F}_t$ consists of the following three components:

\textit{Players:} All MGs participating in the P2P market, denoted as MG 1, MG 2, ..., MG $N^r$.

\textit{Strategies}: The bidding quantity $P^{\rm ex}_{t,r}$ submitted by each MG, given that the trading price $\lambda_t$ is announced by the P2P operator.

\textit{Payoff Function}: The cost function of each MG in the energy market, which is influenced by the P2P trading price and its own bidding quantity, denoted as $f_{t, r}\left(P_{t, r}^{\mathrm{EX}}, \lambda_{t}\right)$.

Next, the equilibrium conditions of the proposed Nash game are mathematically formulated. Specifically, the game equilibrium must satisfy the following four conditions:

\textit{1) Price feasibility condition:}
\begin{equation} \label{eq:price_bound}
    \lambda_t^{\text{FiT}} \leq \lambda_t^{\text{P2P}} \leq \lambda_t^{\text{ToU}}.
\end{equation}

\textit{2) Best-response condition:}
\begin{equation} \label{eq:best_response}
  P_{t,r}^{\text{EX},*} = \underset{P_{t,r}^{\text{EX}}}{\arg\min} f_{t,r}\left(P_{t,r}^{\text{EX}}, \lambda_t^{\text{P2P}}\right), \quad \forall r.
\end{equation}

\textit{3) Supply-demand balance condition:}
\begin{equation} \label{eq:supply_demand}
    \sum_{r=1}^{N^r} P_{t,r}^{\rm EX, *} + P_{t}^{\rm grid} = 0.
\end{equation}

\textit{4) Grid trading complementarity condition:}
\begin{equation} \label{eq:grid_trade}
    P_{t}^{\rm grid} \cdot (\lambda_{t}^{\rm P2P} - \lambda_{t}^{\rm FiT}) \cdot (\lambda_{t}^{\rm P2P} - \lambda_{t}^{\rm ToU}) = 0.
\end{equation}

Condition~\eqref{eq:price_bound} ensures that the market-clearing price remains within an acceptable range determined by the existing ToU and FiT tariffs. Condition~\eqref{eq:best_response} characterizes the rational behavior of each MG, which optimally adjusts its trading quantity in response to the announced market price. Condition~\eqref{eq:supply_demand} ensures energy balance within the P2P market, explicitly incorporating energy exchanged with the main grid.

Finally, condition~\eqref{eq:grid_trade} provides a succinct reformulation of the P2P and P2G relationship introduced earlier in~\eqref{condition}. It implies that if the grid exchange \( P_t^{\rm grid} \) is non-zero, the market-clearing price must exactly reach either the ToU or FiT boundary. However, it should be noted that while~\eqref{condition} explicitly specifies the direction of energy exchange with the grid, condition~\eqref{eq:grid_trade} alone does not fully capture this directional information. Therefore, condition~\eqref{eq:grid_trade} should be interpreted in conjunction with the additional explicit constraints provided in ~\eqref{condition} to fully specify whether energy is imported from or exported to the grid.

The theoretical proof for the existence and uniqueness of this equilibrium will be provided in subsection~\ref{Equilibrium analysis}.

\subsection{Online optimization approach for individual MGs}\label{individual_model}

In each trading interval, MGs determine their best response to the trading price based on their optimization strategies. The internal cost function of each MG, denoted as $f_{t,r}$ in~\eqref{eq:best_response}, significantly influences its operational performance.

\begin{figure}[htbp]
  \footnotesize\rmfamily   \setlength{\abovecaptionskip}{-0.1cm}  
    \setlength{\belowcaptionskip}{-0.1cm} 
  \begin{center}  \includegraphics[width=0.5\columnwidth]{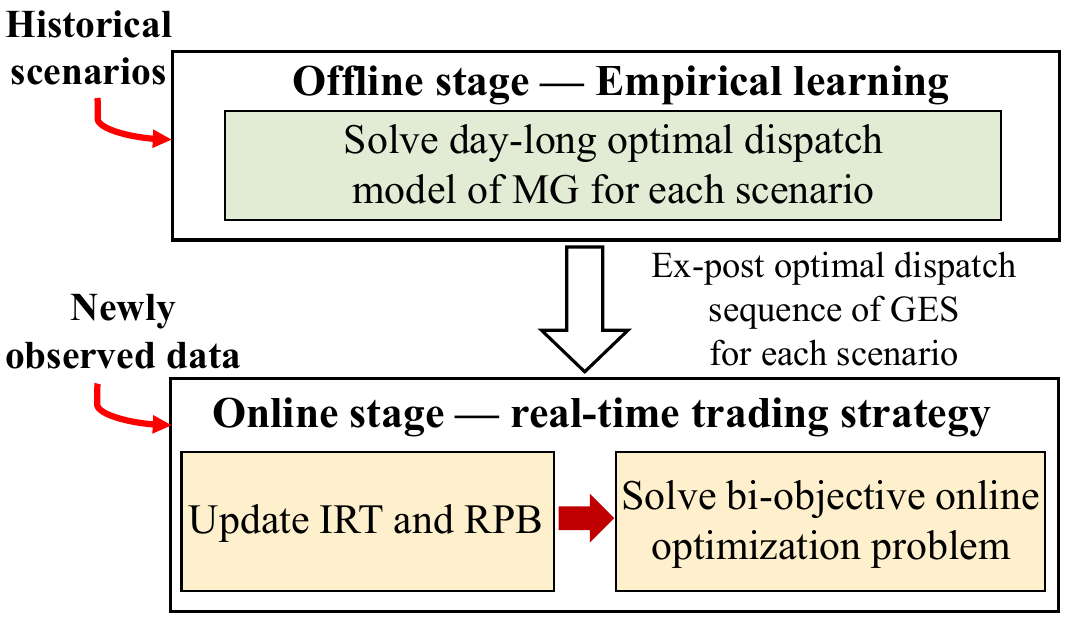}
     \caption{\rmfamily Prediction-free two-stage DDOO framework for MG.}\label{scheme2}
  \end{center}
  \vspace{-2.5em}
\end{figure}

In real-time P2P trading, both trading prices and RES generation within MGs are highly uncertain, making accurate prediction challenging. Moreover, the presence of time-coupled entities such as ES increases the risk of traditional online optimization falling into local optima. To address these challenges, this subsection proposes a two-stage DDOO framework, as illustrated in Fig.~\ref{scheme2}. In the offline stage, historical scenarios are utilized to derive optimal dispatch sequences for generalized energy storage (GES). In the online stage, the IRT and RPB are dynamically updated based on both offline results and newly observed data, and a bi-objective online optimization problem is solved in real-time to provide the MG's real-time trading strategy.


\subsubsection{Offline stage: empirical learning}

Compared to predictions, historical data are much more accessible. Therefore, our approach is to learn from a large number of historical dispatch scenarios to provide valuable experiences for online decision-making. Typically, MG dispatch follows a daily cycle. Specifically, we use a set of historical trading price scenarios and historical net load scenarios (defined as load power minus RES generation), denoted as:
\begin{equation}\label{offline1}
\{\ell_{t,s},\ \lambda_{t,s}^{\text{P2P}}\}_{t=1}^{T}\text{,}\ s\in\{1\text{,}2\text{,}\cdotp\cdotp\cdotp\text{,}S\}\text{,}
\end{equation}

For each scenario, we solve a day-long optimal dispatch model of the MG. A typical MG consists of RES, DG, ES, flexible loads, and conventional loads, and is connected to the utility grid via a tie-line. Many flexible loads, such as thermostatically controlled loads and electric vehicles, exhibit characteristics similar to ES, leading to the concept of “virtual energy storage”~\cite{niromandfam2020virtual}. To enable a unified modeling approach, both ES and virtual energy storage are incorporated within a GES framework. The transformation from physical loads to the GES model is detailed in~\cite{qi2023chance}. The corresponding dispatch model is formulated as follows.

Objective function:
\begin{subequations}\label{all_object}
\begin{align}
    \label{objective}
    &\min\ \sum_{t\in\boldsymbol{\Omega}_T}\left(C_t^\mathrm{G}+C_t^\mathrm{EX}+C_t^\mathrm{DG}\right)\\
   \label{GES cost}  &C_t^\mathrm{G}=(c_{\mathrm{d}}^\mathrm{G}P_{t}^\mathrm{G,d}+c_{\mathrm{c}}^\mathrm{G}P_{t}^\mathrm{G,c})\Delta t\\
    \label{grid_cost}  &C_{t}^{\mathrm{EX}}=\lambda_{t}^{\text{P2P}}P_{t}^{\mathrm{EX}}\Delta t\\
   \label{DG cost}  &C_t^\mathrm{DG}=aP_{t}^\mathrm{DG}\Delta t
\end{align}
\end{subequations}

Constraints: $\forall t\in\boldsymbol{\Omega}_T,\forall i\in\boldsymbol{\Omega}_B$
\begin{equation}\label{DG_output}
   \underline{P}^{\mathrm{DG}}\leq P_{t}^{\mathrm{DG}}\leq\overline{P}^{\mathrm{DG}}
\end{equation}
\begin{equation}\label{GES_leq}
    0\leq P_{{t}}^{{\mathrm{G,c}}},\quad0\leq P_{{t}}^{{\mathrm{G,d}}}
\end{equation}
\begin{equation}\label{GES_output}
\mathbb{P}\left(P_{t}^\mathrm{G,c}\leq\overline{P}_{t}^\mathrm{G,c}\right)\geq1-\varepsilon,\ \mathbb{P}\left(P_{t}^\mathrm{G,d}\leq\overline{P}_{t}^\mathrm{G,d}\right)\geq1-\varepsilon
\end{equation}
\begin{equation}\label{GES_SOC}
\begin{gathered}
\begin{aligned}
SoC_{t+1} =&(1-\mathcal{E})SoC_{t}+\eta^{\mathrm{c}}P_{t}^\mathrm{G,c}\Delta t/E \\
&-P_{t}^\mathrm{G,d}\Delta t/(\eta^{\mathrm{d}}E)+\pi_{t}
\end{aligned}
\end{gathered}
\end{equation}
\begin{equation}\label{GES_SOC_limit}
\mathbb{P}\left(\underline{SoC}_{t}\leq SoC_{t}\leq\overline{SoC}_{t}\right)\geq1-\varepsilon
\end{equation}
\begin{equation}\label{GES_SOC_cycle}
SoC_{T}=SoC_{0}
\end{equation}
\begin{equation}\label{P_balance}
\begin{aligned}
P_{t}^{\mathrm{EX}} = P_{t}^{\mathrm{L}}-P_{t}^{\mathrm{R}}+P_{t}^\mathrm{G,c}-P_{t}^\mathrm{G,d}-P_{t}^\mathrm{DG}
\end{aligned}
\end{equation}
\begin{equation}\label{P_flow}
\begin{aligned}
P_{b+1,t}^{\mathrm{PF}}=&P_{b,t}^{\mathrm{PF}}-P_{b+1,0,t}^{\mathrm{PF}}-P_{i,t}^{\mathrm{L}}+P_{i,t}^{\mathrm{R}}+P_{i,t}^{\mathrm{DG}}\\&+P_{i,t}^{\mathrm{G,d}}-P_{i,t}^{\mathrm{G,c}}, b\in Br(i)
\end{aligned}
\end{equation}
\begin{equation}\label{Q_flow}
\begin{aligned}
Q_{b+1,t}^\mathrm{PF}=&Q_{b,t}^\mathrm{PF}-Q_{b+1,0,t}^\mathrm{PF}-Q_{i,t}^\mathrm{L}\\&+Q_{i,t}^{\mathrm{G,d}}-Q_{i,t}^{\mathrm{G,c}},\ b\in Br(i)
\end{aligned}
\end{equation}
\begin{equation}\label{V_def}
V_{i+1,t}^{\mathrm{BUS}}=V_{i,t}^{\mathrm{BUS}}-(R_bP_{b,t}^{\mathrm{PF}}+X_bQ_{b,t}^{\mathrm{PF}})/V^{\mathrm{B}},\ b\in Br(i,i+1)
\end{equation}
\begin{equation}\label{V_limit}
1-\overline{V}^{\mathrm{BUS}}\leq V_{i,t}^{\mathrm{BUS}}\leq1+\overline{V}^{\mathrm{BUS}}
\end{equation}

The objective function in ~\eqref{objective} minimizes the overall operating cost of the MG, which includes the GES incentive cost~\eqref{GES cost}, the energy trading cost~\eqref{grid_cost}, and the generation cost of DG~\eqref{DG cost}. The DG's output capacity is constrained by~\eqref{DG_output}, while~\eqref{GES_leq}--\eqref{GES_output} restrict the charging and discharging power of the GES. Given that sufficient conditions are satisfied, the complementary constraint for charging and discharging can be relaxed, as discussed in~\cite{li2015sufficient}. Constraint~\eqref{GES_SOC} captures the relationship between charging/discharging power, SoC, and additional energy input from baseline consumption.
The time-varying upper and lower bounds on SoC are defined by~\eqref{GES_SOC_limit}. Both the power and SoC limits of GES are time-dependent and uncertain, and can be estimated through data-driven approaches such as load decomposition and parameter identification~\cite{qi2020smart}. Chance constraints~\eqref{GES_output} and~\eqref{GES_SOC_limit} allow for adjustable reliability levels to accommodate different risk preferences~\cite{baker2019joint}. Additionally,~\eqref{GES_SOC_cycle} enforces an energy-neutral condition over the scheduling horizon to maintain long-term GES sustainability. Constraint~\eqref{P_balance} represents the power balance. Power flow in the radial MG is modeled using the $\textit{DistFlow}$ formulation~\cite{li2019temporally}, as specified in~\eqref{P_flow}--\eqref{V_limit}.

The time-varying and stochastic nature of the power and SoC limits of GES makes the chance constraints~\eqref{GES_output} and~\eqref{GES_SOC_limit} challenging to solve directly. To address this, we adopt a deterministic and tractable reformulation~\cite{qi2023chance}. We apply the standard reformulation proposed in~\cite{vrakopoulou2017chance}, which leads to:
\begin{subequations}\label{reformation}
\begin{align}
    \label{Pc_reform}
&P_{t}^{\mathrm{G,c}}\leq\mu_{\overline{P}_{t}^{\mathrm{G},c}}-F_{\overline{P}_{t}^{\mathrm{G},c}}^{-1}(1-\varepsilon)\gamma_{\overline{P}_{t}^{\mathrm{G,c}}}\\
   \label{Pd_reform} &P_{t}^{\mathrm{G,d}}\leq\mu_{\overline{P}_{t}^{\mathrm{G,d}}}-F_{\overline{P}_{t}^{\mathrm{G,d}}}^{-1}(1-\varepsilon)\gamma_{\overline{P}_{t}^{\mathrm{G,d}}}\\
    \label{SOC_up_reform}  &SoC_{t}\leq\mu_{\overline{SoC}_{t}}-F_{\overline{SoC}_{t}}^{-1}(1-\varepsilon)\gamma_{\overline{SoC}_{t}}\\
   \label{SOC_down_reform}  &SoC_{t}\geq\mu_{\underline{SoC}_{t}}-F_{\underline{SoC}_{t}}^{-1}(1-\varepsilon)\gamma_{\underline{SoC}_{t}}
\end{align}
\end{subequations}
The normalized inverse cumulative distribution function $F^{-1}$ can be obtained by Monte Carlo sampling of any
kind of distribution (e.g., normal distribution, beta distribution)~\cite{homem2014monte}. $\mu$ and $\gamma$ denote the mean and standard deviation, respectively. Thus, the dispatch model becomes convex programming that can be efficiently solved using commercial solvers.

After solving the day-long optimal dispatch model of the MG for each scenario in~\eqref{offline1}, we obtain $S$ offline optimal SoC sequences for GES: $SoC_s=\{SoC_{t,s}\}_{t=1}^T$.

\subsubsection{Online stage: real-time trading strategy}
\begin{algorithm}[htbp]\label{algorithm1}
\caption{Prediction-Free Two-Stage DDOO}
\SetAlgoLined
\SetKwInOut{Input}{Input}
\SetKwInOut{Output}{Output}

\textbf{Stage 1: Empirical Learning}\\
\Input{Historical scenarios $\{\ell_{t,s}, \lambda_{t,s}^{\text{P2P}}\}$.}

\For{$s=1,2,\dots,S$}{
    Solve day-long dispatch~\eqref{all_object}--\eqref{GES_leq},~\eqref{GES_SOC_cycle}--\eqref{reformation} and obtain optimal SoC sequences $\{SoC_{t,s}^*\}_{t=1}^{T}$.\\[2pt]
}
\Output{Optimal SoC sequences $\{SoC_{t,s}^*\}$.}

\vspace{0.5em}
\textbf{Stage 2: Real-Time Trading Strategy}\\
\Input{Offline optimal SoC sequences $\{SoC_{t,s}^*\}$.}

\For{$t=1,2,\dots,T$}{
    Observe uncertainties realization $\ell_{[t]},\lambda_{[t]}^{\text{P2P}}$ and calculate weights $\rho_t^s$ by \eqref{online4} and \eqref{online4_}.\\[2pt]
    Update IRT by \eqref{online3} and RPB by \eqref{DAP}.\\[2pt]
    Solve online optimization problem \eqref{eq:modified_obj} and implement real-time dispatch decisions.\\[2pt]
}
\Output{Real-time dispatch decisions.}
\end{algorithm}

In the P2P market, each MG makes a decision every five minutes based on the current system state and market price. Before each decision, both the IRT and RPB are dynamically updated. 

The IRT is constructed as a weighted combination of the optimal SoC sequences obtained in the offline stage. Specifically, if a historical scenario closely resembles the current day, its corresponding optimal SoC sequence is assigned a higher weight in the IRT.

Let vectors $\ell_{[t]}$ and $\lambda_{[t]}^{\text{P2P}}$ denote the realized uncertainties in net load and price from the start of the operating day up to period $t$.

\begin{equation}\label{online1}
\ell_{[t]}=[\ell_{1}\text{,}\ \cdotp\cdotp\cdotp\text{,}\ \ell_{t}]\text{,}\ \lambda_{[t]}^{\text{P2P}}=[\lambda_{1}^{\text{P2P}}\text{,}\ \cdotp\cdotp\cdotp\text{,}\ \lambda_{t}^{\text{P2P}}].
\end{equation}

Define the vectors representing the \textit{s}th historical scenario:
\begin{equation}\label{online2}
\ell_{[t],s}=[\ell_{1,s}\text{,}\ \cdotp\cdotp\cdotp\text{,}\ \ell_{t,s}]\text{,}\ \lambda_{[t],s}^{\text{P2P}}=[\lambda_{1,s}^{\text{P2P}}\text{,}\ \cdotp\cdotp\cdotp\text{,}\ \lambda_{t,s}^{\text{P2P}}].
\end{equation}

The similarity between $\ell_{[t]}$ and $\ell_{[t],s}$ is quantified using the Euclidean distance $||\ell_{[t]} - \ell_{[t],s}||_2$. Based on this, the IRT is sequentially updated as a weighted average of the offline optimal SoC sequences. Let $SoC_t^{\mathrm{ref}}$ denote the IRT value at period $t$, with the update rule given by:
\begin{equation}\label{online3}
SoC_t^{\mathrm{ref}}=\sum_{s=1}^S\rho_t^sSoC_{t,s}\text{,}
\end{equation}
$\rho_t^{s}$ denotes the weight assigned to the \textit{s}th historical scenario, reflecting its similarity to the current operating day. It is computed for each period $t$ using the Nadaraya-Watson kernel regression method~\cite{bierens1988nadaraya}:
\begin{equation}\label{online4}
\rho_{t}^{s}=\frac{K_{t}(\ell_{[t]},\ell_{[t],s})K_{t}(\lambda_{[t]}^{\text{P2P}},\lambda_{[t],s}^{\text{P2P}})}{\sum_{{s^{\prime}=1}}^{S}[K_{t}(\ell_{[t]},\ell_{{[t],s^{\prime}}})K_{t}(\lambda_{[t]}^{\text{P2P}},\lambda_{{[t],s^{\prime}}}^{\text{P2P}})]}\text{,}
\end{equation}
where with $x$ and $y$ as the input vectors, 
\begin{equation}\label{online4_}
K_{t}(x,y)=e^{{-\frac{(\|x-y\|_{2})^{2}}{t\tau}}}.
\end{equation}
is defined as the Gaussian kernel function; $\tau$ is the bandwidth parameter. The weights satisfy $\sum_s \rho_t^s = 1$. As such, $\{\rho_t^s\}_{s=1}^S$ forms a discrete probability distribution, and the IRT can be interpreted as the conditional expectation of the $S$ offline optimal sequences over the operating day. The dynamically updated IRT provides the most applicable historical insights for online operation.

The RPB is a real-time estimate of the average price for the current day, based on both historical price data and the prices observed thus far. It serves as a benchmark for real-time charging and discharging decisions, enabling better utilization of GES for peak-valley arbitrage. Specifically, the RPB is defined as follows:

\begin{equation}\label{DAP}
\overline{\lambda}_t^{\text{P2P}}=\sum_{s=1}^S\frac{K_{t}(\lambda_{[t]}^{\text{P2P}},\lambda_{[t],s}^{\text{P2P}})}{\sum_{{s^{\prime}=1}}^{S}[K_{t}(\lambda_{[t]}^{\text{P2P}},\lambda_{{[t],s^{\prime}}}^{\text{P2P}})]}\overline{\lambda}_{t,s}^{\text{P2P}}\text{,}
\end{equation}
where $\overline{\lambda}_t$ denotes the RPB, and $\overline{\lambda}_{t,s}^{\text{P2P}}$ represents the average price of the \textit{s}th scenario. Similar to the IRT, the RPB can be interpreted as the conditional expectation of the average prices across the \textit{S} scenarios for the operating day. Without the RPB, the GES tends to discharge in order to reduce the immediate cost. To address this, we incorporate the RPB into the formulation by modifying $C_t^{\mathrm{G}}$ in~\eqref{GES cost} as follows:
\begin{equation}\label{revise_GES}
\hat{C}_t^{\mathrm{G}}=((c_{\mathrm{d}}^{\mathrm{G}}+\overline{\lambda}_t^{\text{P2P}})P_{\mathrm{d},t}^{\mathrm{G}}+(c_{\mathrm{c}}^{\mathrm{G}}-\overline{\lambda}_t^{\text{P2P}})P_{\mathrm{c},t}^{\mathrm{G}})\Delta t\text{,}
\vspace{-0.1cm}
\end{equation}

In fact, the immediate operating cost of the MG at time $t$, denoted as $C_t^{\mathrm{all}}$, can be formulated as equation~\eqref{original_cost}. After incorporating the IRT and RPB, the bi-objective online optimization problem at time $t$ is given in ~\eqref{eq:modified_obj}.

\begin{equation}\label{original_cost}
C_t^{\mathrm{all}}=C_t^\mathrm{G}+C_t^\mathrm{EX}+C_t^\mathrm{DG}
\end{equation}
\begin{align}\label{eq:modified_obj}
\min~&f_t=\hat{C}_t^\mathrm{G}+C_t^\mathrm{EX}+C_t^\mathrm{DG}+\varphi(SoC_t-SoC_t^\mathrm{ref})^2  \\
&\mathrm{s.t.~constraints~\eqref{DG_output}\text{--}\eqref{GES_leq},~\eqref{GES_SOC},~\eqref{P_balance}\text{--}\eqref{reformation}} \nonumber
\end{align}

The quadratic penalty term $\varphi(SoC_t-SoC_t^\mathrm{ref})^2$, where $\varphi$ is a weighting coefficient, enforces tracking of the IRT, thereby embedding strategic long-term guidance into real-time decision-making. By combining these elements, the proposed DDOO framework effectively integrates short-term market economics with long-term operational optimality. Constraint~\eqref{GES_SOC_cycle} is omitted since~\eqref{eq:modified_obj} performs local optimization and is not suited for handling inter-temporal constraints. Nevertheless, effective tracking of the IRT can still help maintain long-term energy sustainability for the GES.

\begin{remark}
{\color{blue}
It is worth noting that the historical dataset used for generating the IRT is not static. In practice, the dataset can be updated in a rolling manner; for example, it can be updated day by day. At the end of each operating day, we solve the day-long optimal dispatch problem (the convex model defined by~\eqref{all_object}--\eqref{GES_leq} and~\eqref{GES_SOC_cycle}--\eqref{reformation}) to obtain the hindsight-optimal SoC sequence for that day. The scenario of that day (i.e., the net load and P2P price) together with its corresponding optimal SoC sequence is then added to the historical dataset. From the next day onward, this new record is available when computing the IRT according to~\eqref{online3}--\eqref{online4_}. Therefore, the dataset evolves dynamically, and the IRT will progressively reflect new data patterns. This incremental update incurs very little computational burden: solving one convex day-long model is fast, and the kernel-based computation in~\eqref{online3}--\eqref{online4_} is efficient even if the dataset covers several years. 
To prevent the dataset from infinite growth, a sliding window of recent years can be kept.}
\end{remark}

\begin{remark}
{\color{blue}
This remark clarifies the assumptions on historical data availability, microgrid observability, and the adaptability of DDOO under limited or noisy data conditions. For historical data availability, the framework requires past $S$ days of aggregated net load trajectories and P2P market prices, each recorded at a 5-minute resolution. Only aggregated net load is needed, which can be derived from tie-line power together with the outputs of GES and DG units, without requiring individual user-level data. For real-time operation, each microgrid is assumed to have at least the following observability: (1) real-time power output data of local assets, including RES, battery storage, and DG units; (2) tie-line power exchanged with the external grid; and (3) P2P trading prices issued at each trading interval. If power flow constraints are to be satisfied, real-time active and reactive power injections at each node are also required. Furthermore, to fully leverage flexible loads in demand response, when they are modeled as VES, their real-time power consumption (e.g., electric vehicle charging stations, building loads) should be observable. When only limited historical data are available initially (e.g., several tens of days), scenario generation methods such as attention-based conditional generative adversarial networks (GAN)~\cite{li2024long} or knowledge-integrated GAN~\cite{fu2025knowledge} can be used to enrich the dataset; thereafter, the dataset will be continuously expanded by rolling updates as described in Remark~2. In the presence of noise data due to sensor inaccuracies, communication interference, or data collection disturbances, noise-reduction techniques can be applied. Classical filters (e.g., Kalman or particle filters) and smoothing methods (e.g., exponentially weighted moving average) can effectively reduce the impact of noise and provide more reliable signals.}
\end{remark}

\subsection{Equilibrium analysis and solution algorithm}\label{Equilibrium analysis}

\begin{algorithm}[htbp]
\caption{ASSA}\label{algorithm_ASSA}
\SetAlgoLined
\KwIn{{\color{blue}Initial price $\lambda_{0,t}^{\mathrm{P2P}}=\lambda_{t-1}^{\mathrm{P2P,*}}$}, initial step size $\sigma$, convergence tolerance $\delta$;}
\KwOut{Equilibrium price $\lambda_t^{\mathrm{P2P,*}}$;}

Set iteration index $k \leftarrow 0$\;

\Repeat{convergence or boundary reached}{
    Each MG computes best-response by~\eqref{eq:optimal_response_definition}, and P2P operator calculates imbalance: $F_{k,t}=\sum_{r=1}^{N^r} P_{k,t,r}^{\mathrm{EX,*}}$\;

    \uIf{$|F_{k,t}| \leq \delta$}{
        Equilibrium achieved; set $\lambda_t^{\mathrm{P2P,*}} = \lambda_{k,t}^{\mathrm{P2P}}$; Break\;
    }
    \Else{
        \If{$k \geq 1$ and $F_{k,t}\cdot F_{k-1,t}<0$}{
            Update step size: $\sigma \leftarrow \sigma/2$\;
        }
        Update price: $\lambda_{k+1,t}^{\mathrm{P2P}}\leftarrow\lambda_{k,t}^{\mathrm{P2P}}+\sigma\cdot F_{k,t}$\;

        \uIf{$\lambda_{k+1,t}^{\mathrm{P2P}}<\lambda_t^{\mathrm{FiT}}$}{
            Set boundary price $\lambda_t^{\mathrm{P2P,*}}=\lambda_t^{\mathrm{FiT}}$; Break\;
        }
        \uElseIf{$\lambda_{k+1,t}^{\mathrm{P2P}}>\lambda_t^{\mathrm{ToU}}$}{
            Set boundary price $\lambda_t^{\mathrm{P2P,*}}=\lambda_t^{\mathrm{ToU}}$; Break\;
        }
        \Else{
            $k\leftarrow k+1$\;
        }
    }
}
\Return{$\lambda_t^{\mathrm{P2P,*}}$}
\end{algorithm}

In this subsection, the equilibrium of the non-cooperative Nash game model presented in subsection~\ref{game} is analyzed. Before discussing the existence and uniqueness of the equilibrium, we first present a key intermediate result regarding the monotonic behavior of the MGs' best-response functions with respect to the market price.

\begin{proposition}[Monotonicity of MGs' Best-Response Functions]\label{Proposition1}
Given any microgrid $r$ and trading interval $t$, define the internal cost function explicitly as:
\begin{equation}\label{eq:original_objective}
f_{t,r}(P_{t,r}^{\mathrm{EX}},\lambda_t^{\mathrm{P2P}})=\hat{C}_t^{\mathrm{G}}+C_t^{\mathrm{EX}}+C_t^{\mathrm{DG}}+\varphi(SoC_t - SoC_t^{\mathrm{ref}})^2.
\end{equation}
Then, the optimal response function 
\begin{equation}\label{eq:optimal_response_definition}
P_{t,r}^{\mathrm{EX,*}}(\lambda_t^{\mathrm{P2P}})=\arg\min_{P_{t,r}^{\mathrm{EX}}}f_{t,r}(P_{t,r}^{\mathrm{EX}},\lambda_t^{\mathrm{P2P}}),
\end{equation}
is continuous and monotonically non-increasing with respect to the market-clearing price $\lambda_t^{\mathrm{P2P}}$.
\end{proposition}

A rigorous mathematical proof of Proposition~\ref{Proposition1} is provided in Appendix~A. This monotonicity property is crucial, as it ensures that the aggregate supply-demand function is monotonically decreasing, laying the foundation for the subsequent proof of equilibrium existence and uniqueness.

We now present a more general proposition regarding the existence and uniqueness of the equilibrium for the non-cooperative Nash game $\mathcal{F}_t$ formulated in subsection~\ref{game_formulation}.

\begin{proposition}[Existence and Uniqueness of Nash Equilibrium]\label{Proposition2}
For the Nash game $\mathcal{F}_t$ defined in subsection~\ref{game_formulation}, if each MG's optimal response function $P_{t,r}^{\mathrm{EX,*}}(\lambda_t^{\mathrm{P2P}})$ is continuous and monotonically non-increasing with respect to the market-clearing price $\lambda_t^{\mathrm{P2P}}$, then there exists a unique equilibrium satisfying conditions~\eqref{eq:price_bound}--\eqref{eq:grid_trade}.
\end{proposition}

Proposition~\ref{Proposition2} demonstrates the generality and robustness of the proposed Nash game formulation. Notably, the existence and uniqueness of the equilibrium do not require the MGs' cost functions $f_{t,r}$ to follow the exact form presented in this paper. Rather, any form of continuous and monotonically non-increasing optimal response function is sufficient to guarantee a unique equilibrium. This property aligns closely with economic intuition: as the market-clearing price rises, rational market participants naturally tend to decrease their energy purchases. Therefore, this general property reinforces the applicability and flexibility of our modeling framework across a wide range of practical market scenarios. A detailed proof of Proposition~\ref{Proposition2} is provided in Appendix~B.

In order to efficiently identify the equilibrium price in the real-time P2P market, the ASSA is proposed. The key idea is that the P2P operator iteratively announces a trading price, receives the best-response quantities from MGs, and adaptively updates the price based on the observed supply-demand imbalance. An essential feature of ASSA is its adaptive adjustment of the step-size, effectively preventing oscillations near the equilibrium point and ensuring rapid convergence. Additionally, the algorithm fully preserves the privacy of MGs, as each MG independently computes its best-response locally and only submits the bid quantity to the P2P operator. Moreover, ASSA also supports distributed and parallel computation, with each MG solving only a simple quadratic optimization problem, thereby achieving high computational efficiency. The complete procedure is detailed in Algorithm~\ref{algorithm_ASSA}.

\begin{proposition}[Finite-step Convergence of ASSA]\label{Proposition3}
Consider the proposed ASSA described in Algorithm~\ref{algorithm_ASSA}. Suppose the aggregate supply-demand function \(F_t(\lambda_t^{\mathrm{P2P}})\) is continuously differentiable and strictly monotonically decreasing. Note that the aggregate supply-demand function in this paper naturally meets this requirement. Then, for any given convergence tolerance \(\delta>0\), ASSA converges to the equilibrium price \(\lambda_t^{\mathrm{P2P,*}}\) within a finite number of iterations.\end{proposition}

The mathematical proof of Proposition~\ref{Proposition3} is provided in Appendix~C. Proposition~\ref{Proposition3} demonstrates the finite-step convergence property of ASSA, making it particularly suitable for real-time implementation.

{\color{blue}
\begin{remark}[Choice of the initial trading price]
In the proposed mechanism, the operator-announced initial trading price serves only as the starting point of the iterative updating process. The final market-clearing price is uniquely determined by the supply–demand relationship and the participants’ cost functions, and is therefore independent of the initial value. The monotonicity of the aggregate supply–demand function $F_t(\lambda_t^{\mathrm{P2P}})$ guarantees the existence of a unique equilibrium, and the ASSA ensures finite-step convergence to this equilibrium. For practical implementation, we adopt the last-interval clearing price as the default initialization. Since adjacent 5-minute clearing prices usually do not change drastically, this choice often places the initial point closer to the new equilibrium compared with selecting FiT or ToU boundaries directly. As a result, the number of iterations required for convergence can be reduced, while the equilibrium outcome remains unaffected by the initial value.
\end{remark}}

{\color{blue}
\begin{remark}[Strategic manipulation]
While the proposed market mechanism aims to ensure efficient and fair market outcomes, we acknowledge that strategic manipulative behaviors could still arise in iterative auctions. Nevertheless, our mechanism inherently mitigates this risk. First, since the P2P operator centrally updates the price and participants submit only quantities, direct price manipulation through price bids is effectively prevented. Second, quantity manipulation is constrained by a ``volume--price trade-off,'' where higher unit prices gained by artificially reducing supply are typically offset by lower cleared volumes. This effect becomes more significant as the number of participants increases, further diluting prosumers' marginal impact on the market price. Third, large-capacity participants have limited incentives to oversupply, as the P2P operator continually lowers prices under excess supply conditions. Finally, whenever supply–demand balance is achieved, the clearing price, bounded between FiT and ToU tariffs, ensures individual rationality and maintains profitability for all participants. Practical safeguards can further enhance market robustness against strategic manipulation, including: (i) enforcing penalties for misreporting to discourage deviations between reported bids and actual delivered quantities; (ii) imposing ramping limits on quantity adjustments between iterations to prevent participants from drastically changing quantities to disturb convergence; and (iii) requiring strict submission deadlines with penalties to prevent delayed reporting intended to interfere with the iterative process.
\end{remark}}

\section{Case studies}~\label{case}
\begin{table}[htbp]
\centering
\footnotesize\rmfamily
\caption{\rmfamily Parameters of the 20 MGs}
\begin{tabular}{cc}
\toprule
\textbf{Parameter} & \textbf{Range} \\ \midrule
Rated wind power capacity & 400–900 kW \\
Rated photovoltaic capacity & 200–400 kW \\
Load capacity & 200–800 kW \\
Rated DG power & 100–250 kW \\
ES capacity / ES duration & 500–1300 kWh / 2–4 hours \\
VES capacity / VES duration & 300–600 kWh / 2–3 hours\\
Cost coefficients for GES & \$0.012–\$0.025/kWh \\
DG generation cost & \$0.12–\$0.19/kWh \\
Grid topology & IEEE 12-bus, 15-bus, 33-bus systems \\ \bottomrule
\end{tabular}
\label{tab:microgrid_parameters}
\end{table}
\subsection{Set-up}
\begin{figure}[htbp]
  \footnotesize\rmfamily   \setlength{\abovecaptionskip}{-0.1cm}  
    \setlength{\belowcaptionskip}{-0.1cm} 
  \begin{center}  \includegraphics[width=0.5\columnwidth]{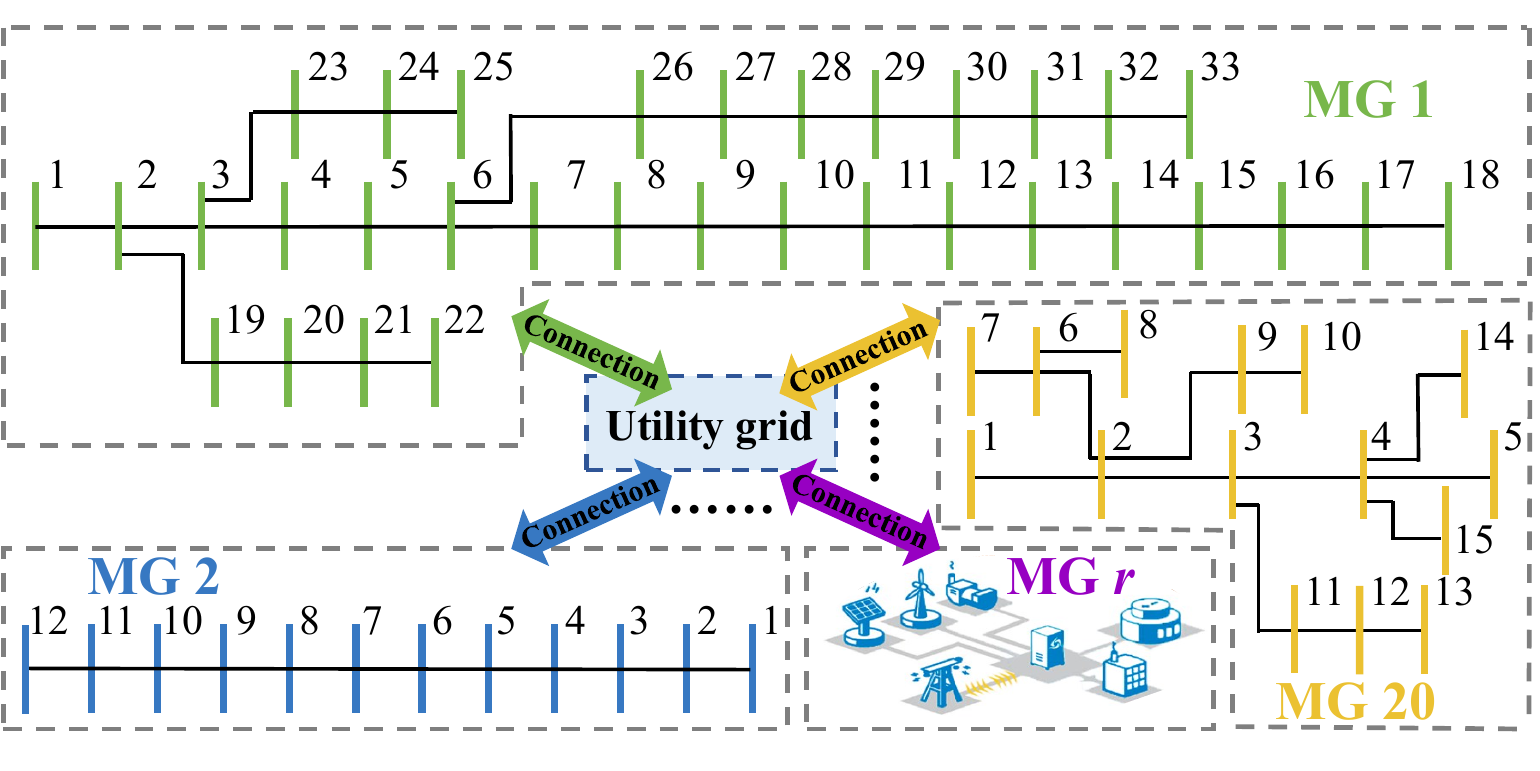}
     \caption{\rmfamily The energy market with the utility grid and 20 MGs. }\label{test_system}
  \end{center}
  \vspace{-2.5em}
\end{figure}

\begin{figure}[htbp]
  \footnotesize\rmfamily   \setlength{\abovecaptionskip}{-0.1cm}  
    \setlength{\belowcaptionskip}{-0.1cm} 
  \begin{center}  \includegraphics[width=0.5\columnwidth]{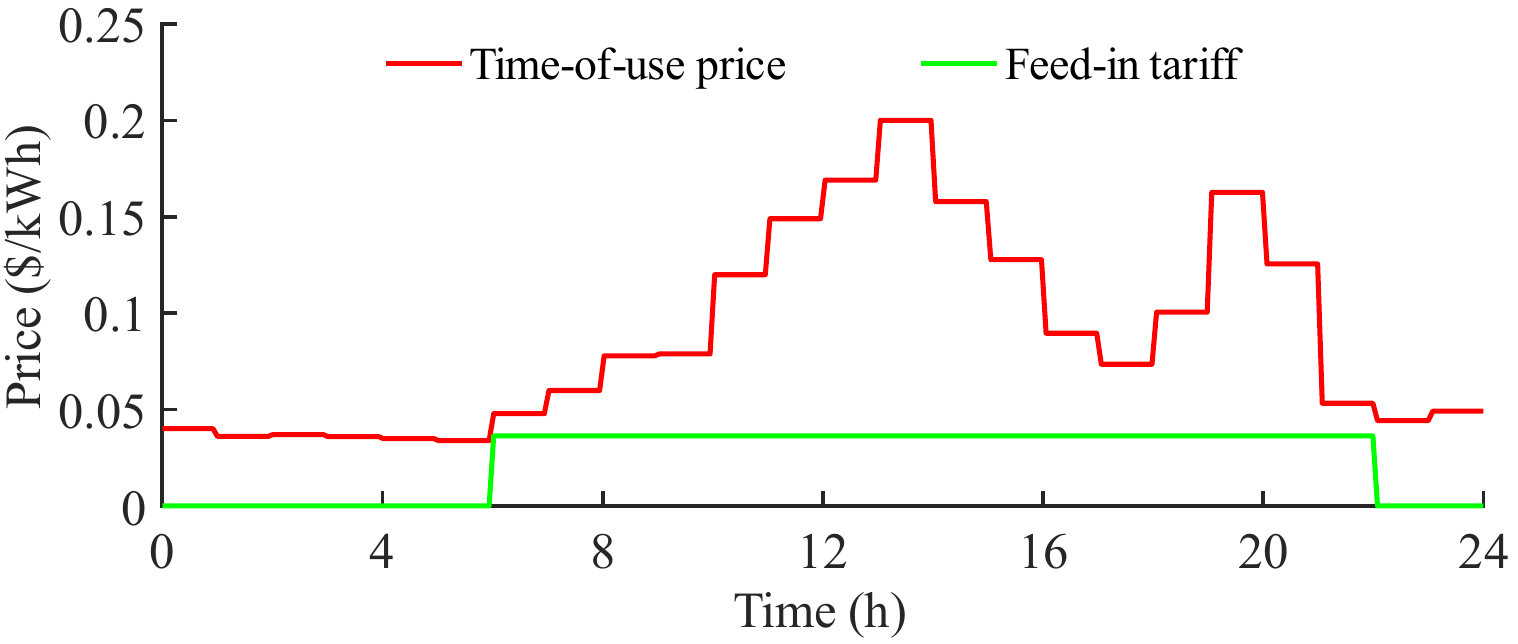}
     \caption{\rmfamily The ToU and FiT of the utility grid.}\label{ToU_FiT}
  \end{center}
  \vspace{-2.5em}
\end{figure}

To verify the effectiveness of the proposed method, an integrated P2P energy market with the utility grid and 20 interconnected MGs as shown in Fig.~\ref{test_system} is studied. The ToU and FiT schemes applied by the utility grid are depicted in Fig.~\ref{ToU_FiT}. 

The parameters of the 20 MGs, randomly generated within the ranges listed in Table~\ref{tab:microgrid_parameters}, are used for the case study. The case study involves continuous P2P trading over 60 days, with trading occurring every five minutes. The voltage at each bus is constrained to remain within a range of 1±0.05 per unit (p.u.). All relevant data, including detailed parameters of each microgrid, 60 days of RES generation and load data, and time-varying SoC bounds for GES, are made publicly available~\cite{huangdata}. The optimization models are implemented in MATLAB using the YALMIP interface, and the problem is solved using the Gurobi 11.0 solver. The programming environment is Intel Core i9-13900HX @ 2.30 GHz with RAM 32GB.

\subsection{The value of P2P energy trading}

We compare two cases: Case I, where P2P trading is conducted using the proposed market mechanism, and Case II, where no P2P trading is employed, and each MG trades independently with the utility grid under ToU and FiT schemes. Through this analysis, the impact of P2P trading on local energy consumption and MG operating costs is evaluated.

\begin{figure*}[htbp]
  \footnotesize\rmfamily   \setlength{\abovecaptionskip}{-0.1cm}  
  \begin{center}  \includegraphics[width=1\columnwidth]{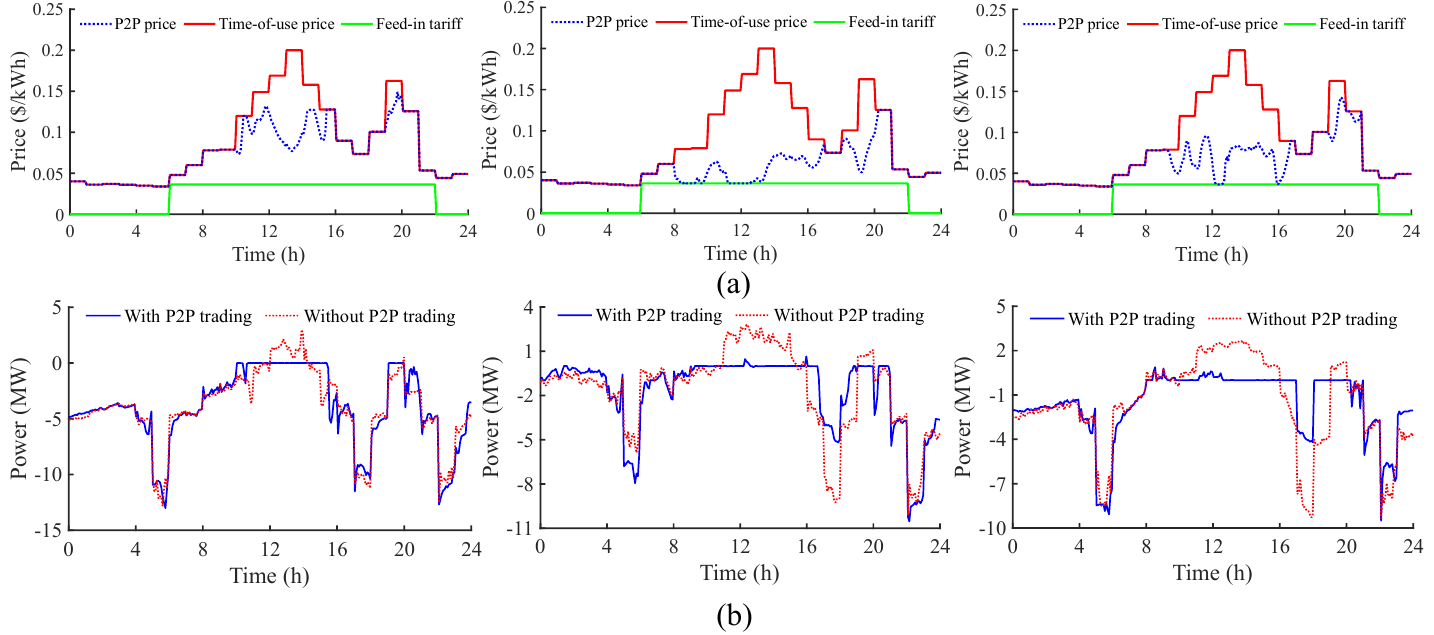}
     \caption{\rmfamily (a) P2P market trading results and (b) power exchange with utility grid (day 11, day 21, and day 31 from left to right).}
     \captionsetup{justification=centering}\label{trading_result}
  \end{center}
  \vspace{-2.5em}
\end{figure*}
\begin{figure*}[htbp]
  \footnotesize\rmfamily   \setlength{\abovecaptionskip}{-0.1cm}  
    \setlength{\belowcaptionskip}{-0.1cm} 
  \begin{center}  \includegraphics[width=1\columnwidth]{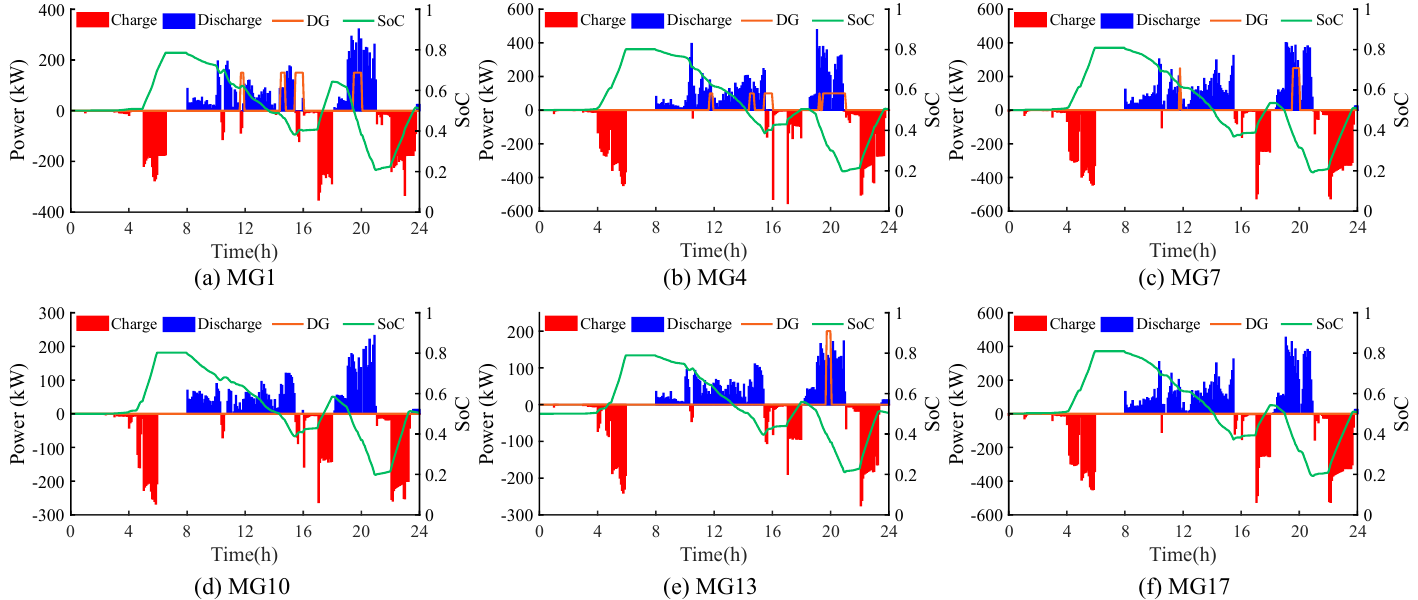}
     \caption{\rmfamily Energy management strategies of several MGs on day 11.}
     \captionsetup{justification=centering}\label{comparison_case}\
  \end{center}
  \vspace{-2.5em}
\end{figure*}

Figure~\ref{trading_result} (a) illustrates the P2P trading results on days 11, 21, and 31. As expected, the P2P clearing price remains between the ToU and FiT tariffs, aligning with the price feasibility condition discussed earlier. The dynamic fluctuation of the P2P price clearly reflects the real-time changes in the energy supply-demand balance within the market. By enabling simultaneous bidding from both buyers and sellers, the improved double auction mechanism promotes market competition and thus enhances the efficiency of resource allocation.

In Figure~\ref{trading_result} (b), the power exchange \( P_t^{\text{grid}} \) between the 20-MG system and the utility grid is presented for the same three days, verifying the previously introduced P2P and P2G relationship. Specifically, when $\lambda_t^{\text{P2P}} = \lambda_t^{\text{ToU}}$, the system imports power from the utility grid. Conversely, when $\lambda_t^{\text{P2P}} = \lambda_t^{\text{FiT}}$, the system exports power to the grid. Notably, when $\lambda_t^{\text{FiT}} < \lambda_t^{\text{P2P}} < \lambda_t^{\text{ToU}}$, the MGs collectively achieve self-sufficiency, balancing local supply and demand without interacting with the utility grid. Compared with the case without P2P trading, the proposed mechanism effectively promotes local energy consumption. This is particularly evident during periods of peak photovoltaic generation (10:00–16:00), where substantial reverse power flow would otherwise occur. By facilitating local self-sufficiency, P2P trading significantly mitigates the impact of RES reverse flows on the utility grid.

Figure~\ref{comparison_case} presents the energy management strategies of several MGs on day 11. Each MG dispatches its DG and GES resources using the proposed DDOO framework to actively participate in the P2P market. Despite the absence of predictions, the MGs successfully perform peak-valley arbitrage in response to the dynamic P2P price signals, effectively reducing their operating costs. Compared with the case without P2P trading, each MG achieves a lower operating cost. Specifically, during self-sufficiency periods, MGs selling surplus energy benefit from increased revenues as the P2P price exceeds the FiT, while MGs purchasing energy reduce their expenses by buying electricity at a P2P price lower than the ToU.

\begin{table}[b!]
\centering
\footnotesize\rmfamily
\caption{\rmfamily Comparison of different cases during the 60-day operation}
\begin{tabular}{c p{1.2cm}<{\centering} p{1.2cm}<{\centering}}
\toprule
\textbf{Metric} & \textbf{Case I} & \textbf{Case II} \\ \midrule
Self-sufficiency period (\%) & 29.86 & 0.01 \\
Reverse power flow period (\%) & 3.96 & 24.51 \\
Average operating cost of each MG (\$) & 14228 & 17609 \\ 
\bottomrule
\end{tabular}
\label{tab:case_comparison}
\end{table}

The comparison of the two cases is summarized in Table~\ref{tab:case_comparison}. Three key metrics are evaluated to quantify the advantages of adopting the proposed P2P trading mechanism.

The self-sufficiency period significantly increases from 0.01\% in Case II to 29.86\% in Case I, demonstrating that the interconnected MG system achieves a perfect local energy balance for approximately 30\% of the time. This result indicates the effectiveness of the proposed mechanism in enhancing local RES utilization. The period of reverse power flow decreases substantially from 24.51\% in Case II to 3.96\% in Case I. This reduction highlights the capability of P2P trading mechanism to mitigate the undesirable reverse energy flows typically caused by excess RES generation, thereby decreasing potential stress and instability on the utility grid. Additionally, the average operating cost for each MG decreases by 19.20\%. This cost reduction is mainly attributed to the more favorable P2P market prices compared to the FiT and ToU tariffs.

{\color{blue}
Overall, the results confirm that the proposed market mechanism effectively facilitates local RES consumption, alleviates the impact of reverse power flows on the utility grid, and significantly reduces the operating costs of MGs. Specifically, the significant increase in self-sufficiency periods highlights that nearly one-third of the time the interconnected system can achieve complete local balance without relying on the external grid, thereby maximizing the utilization of distributed renewable resources. The drastic reduction in reverse power flow periods demonstrates that P2P trading provides an effective means to absorb surplus RES generation locally, which is particularly important in small-scale systems where reverse flows can cause voltage fluctuations and threaten grid stability. Finally, the observed 19.2\% reduction in average MG operating costs shows that the proposed trading mechanism creates tangible economic benefits for both sellers and buyers by establishing prices more favorable than FiT and ToU tariffs. These findings not only validate the superiority and practical effectiveness of the proposed mechanism under realistic operating conditions, but also provide new evidence that P2P market designs can simultaneously enhance renewable integration, improve system reliability, and increase participant profitability, thereby addressing critical gaps left by conventional trading frameworks.}

\begin{figure*}[htbp]
  \footnotesize\rmfamily   \setlength{\abovecaptionskip}{-0.1cm}  
    \setlength{\belowcaptionskip}{-0.1cm} 
  \begin{center}  \includegraphics[width=0.9\columnwidth]{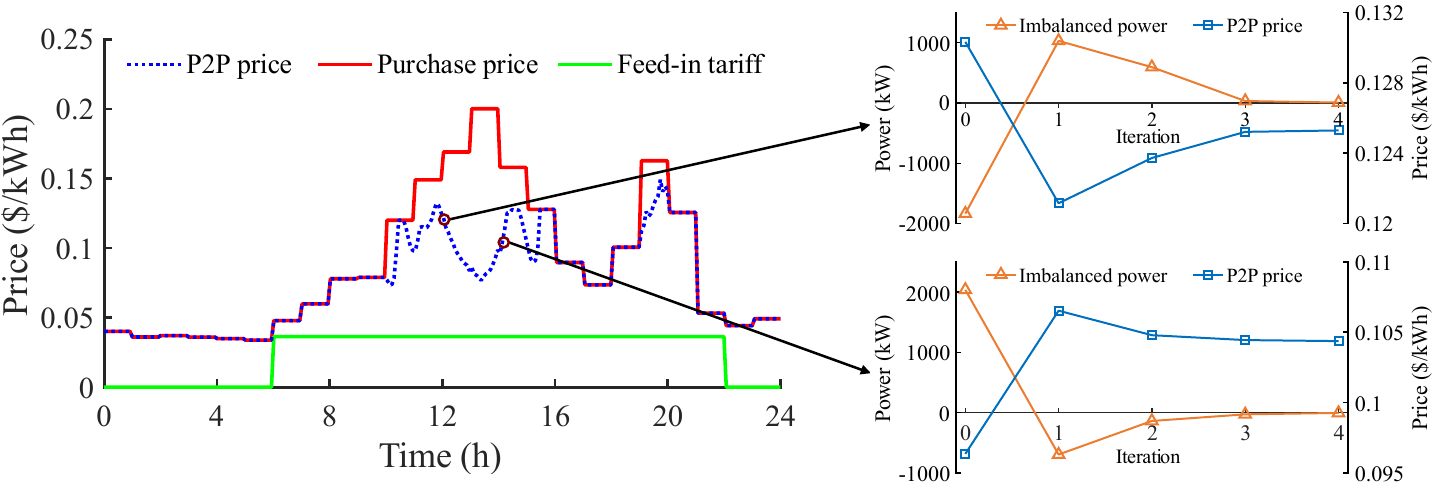}
     \caption{\rmfamily Convergence process of market equilibrium calculation at two time intervals on day 11.}
     \captionsetup{justification=centering}\label{Convergence process}
  \end{center}
  \vspace{-0.5cm}
\end{figure*}

\begin{figure*}[htbp]
  \footnotesize\rmfamily   \setlength{\abovecaptionskip}{-0.1cm}  
    \setlength{\belowcaptionskip}{-0.1cm} 
  \begin{center}  \includegraphics[width=1\columnwidth]{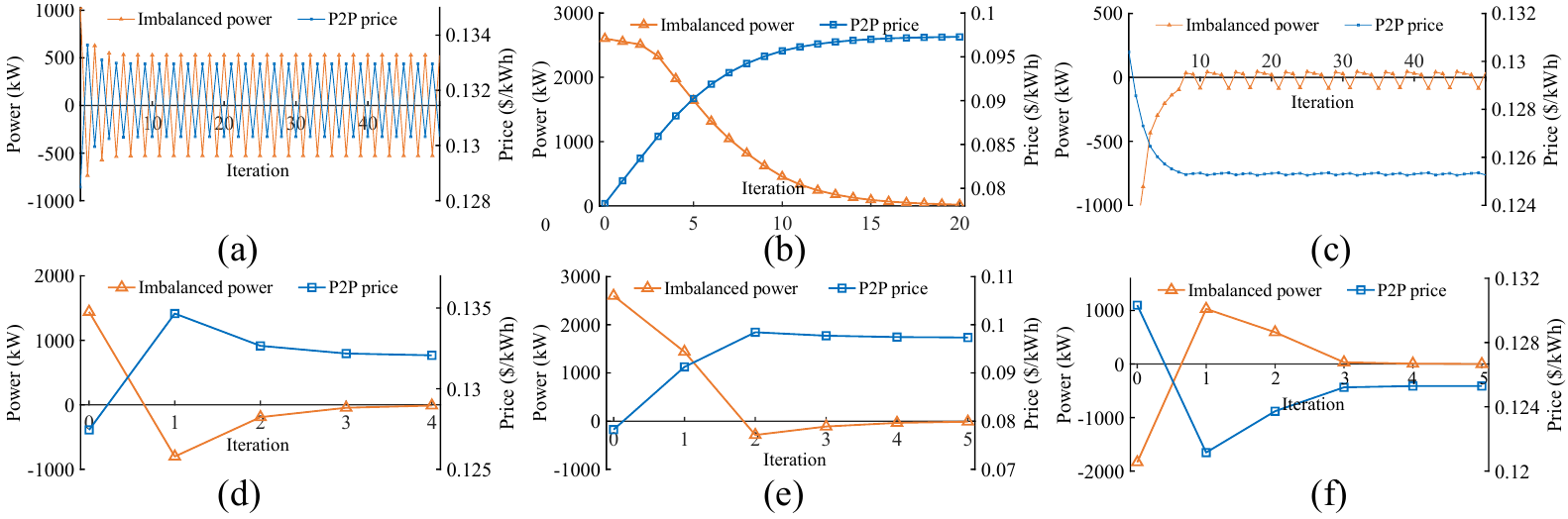}
     \caption{\rmfamily  Convergence comparison between fixed step-size search and ASSA on day 11: (a) Fixed step-size (\(\sigma=5\times10^{-6}\)) at 11:45; (b) Fixed step-size (\(\sigma=1\times10^{-6}\)) at 10:20; (c) Fixed step-size (\(\sigma=1\times10^{-6}\)) at 11:55; (d) ASSA at 11:45; (e) ASSA at 10:20; (f) ASSA at 11:55.}
     \captionsetup{justification=centering}\label{Convergence comparison}
  \end{center}
  \vspace{-0.5cm}
\end{figure*}
\begin{figure*}[htbp]
  \footnotesize\rmfamily   \setlength{\abovecaptionskip}{-0.1cm}  
    \setlength{\belowcaptionskip}{-0.1cm} 
  \begin{center}  \includegraphics[width=1\columnwidth]{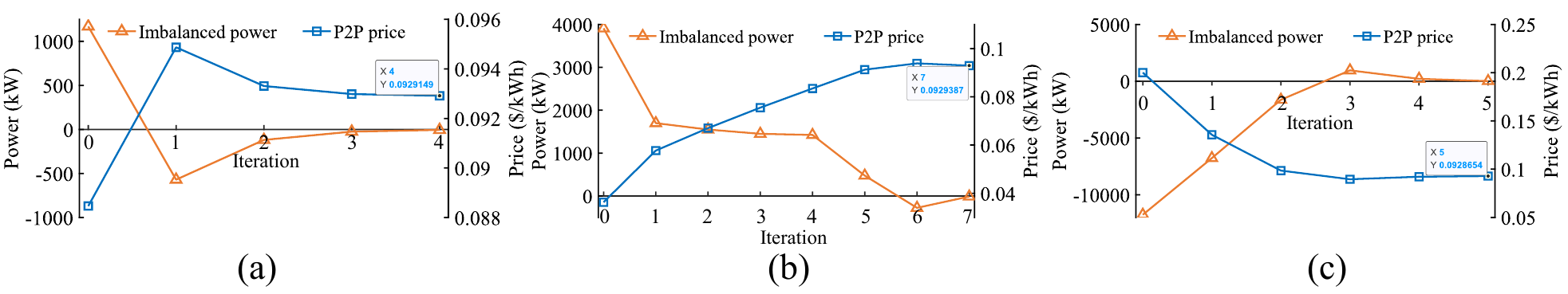}
     \caption{\rmfamily  {\color{blue}Convergence of the iterative process at 13:50 on day 11 under different initial prices: (a) last-interval clearing price, (b) FiT, and (c) ToU.}}
     \captionsetup{justification=centering}\label{initial}
  \end{center}
  \vspace{-1cm}
\end{figure*}
\subsection{Performance evaluation of ASSA}

This subsection analyzes the convergence performance of the proposed ASSA. The initial step size at each time interval is set as \(\sigma=5\times10^{-6}\), and the convergence tolerance is set as \(\delta=5\,\text{kW}\). Table~\ref{tab:convergence_performance} summarizes the average number of iterations and computation time required for reaching equilibrium at each time interval during the 60-day operation.

\begin{table}[htbp]
\centering
\footnotesize\rmfamily
\caption{\rmfamily Convergence performance of ASSA per interval over 60 days}
\begin{tabular}{ccc}
\toprule
\textbf{Period} & \makecell{\textbf{Iterations}\\\textbf{per interval}} & \makecell{\textbf{Computation Time}\\\textbf{per interval (s)}} \\ 
\midrule
Self-sufficiency period & 4.52 & 4.72 \\
Non-self-sufficiency period & 1.02 & 1.06 \\
Overall & 2.07 & 2.16 \\
\bottomrule
\end{tabular}
\label{tab:convergence_performance}
\end{table}

Table~\ref{tab:convergence_performance} shows that the proposed market mechanism significantly reduces the number of computations required by each MG compared to traditional high-dimensional bidding strategies in double auction markets. For typical 20-dimensional and 10-dimensional bid settings, the proposed method achieves speedup factors of approximately 9.68 and 4.84, respectively, as calculated by~\eqref{speed_up}. Furthermore, each MG only solves a simple quadratic optimization problem described by~\eqref{eq:original_objective}, which can be efficiently handled by commercial solvers. The average computational time for continuous 24-hour P2P trading is 622 seconds (with parallel computing among MGs), validating that the proposed ASSA combined with the DDOO framework is well-suited for real-time operations with 5-minute clearing intervals.

Figure~\ref{Convergence process} presents the convergence process during typical self-sufficiency periods. Initially, due to large power imbalances, the first iteration drives the price rapidly towards, or even beyond, the equilibrium point. Subsequently, ASSA adaptively reduces the step size by half, effectively avoiding oscillations and quickly converging to the equilibrium.

The adaptive step-size adjustment in ASSA is critical for ensuring convergence. As shown in Figure~\ref{Convergence comparison}, a fixed step size cannot guarantee convergence at all times. For example, Figure~\ref{Convergence comparison}(a) demonstrates persistent oscillations near the equilibrium point when using a fixed step size. Reducing the step size cannot universally solve this issue, as a smaller step size either results in slow convergence (Figure~\ref{Convergence comparison}(b)) or still fails to eliminate oscillations in some intervals (Figure~\ref{Convergence comparison}(c)). According to the theoretical analysis in~\eqref{step_condition} (Appendix C), the fixed-point iteration convergence theorem states that convergence cannot be achieved when \(\sigma>\frac{2}{|F_t'(\lambda_t^{\mathrm{P2P,*}})|}\). Since the value of \(F_t'(\lambda_t^{\mathrm{P2P,*}})\) is time-varying and unknown to the P2P operator, it is impossible to pre-select a suitable step size for all time intervals. Therefore, the adaptive step-size adjustment in ASSA is essential to guarantee convergence across all time intervals. An initial moderate step size ensures rapid convergence toward the equilibrium, while adaptive adjustments ensure stable convergence thereafter.

{\color{blue}
Finally, we investigate the impact of the choice of the initial price on the convergence of ASSA. Figure~\ref{initial} reports the convergence at 13:50 on day~11 under different inital prices: (a) using the last-interval clearing price as the initial value, which converges in 4 iterations; (b) using FiT as the initial value, which converges in 7 iterations; and (c) using ToU as the initial value, which converges in 5 iterations. In all three cases, the algorithm converges to the same equilibrium price of approximately 0.0929 \$/kWh, demonstrating that the equilibrium outcome is independent of the initial price.}

{\color{blue}
\subsection{Comparison with conventional double auction}

This subsection compares the proposed mechanism with a baseline community market model that adopts a conventional double auction without the ASSA enhancements. Since DDOO is a microgrid operational strategy rather than a market-level mechanism, it is applied uniformly so that the observed differences reflect only the market mechanism design.

The conventional benchmark follows the widely used double auction mechanism adopted in practice by major market operators such as PJM, CAISO, and AEMO. In this setting, each participant submits multiple price–quantity pairs at every trading interval. The P2P operator aggregates all submitted bids, and the clearing price is determined as the value that minimizes the absolute supply–demand imbalance. To capture different levels of bidding granularity, we consider three cases with 5-dimensional, 10-dimensional, and 20-dimensional bidding. For example, in 10-dimensional bidding, each microgrid submits 10 distinct price–quantity pairs at every trading interval.

All mechanisms are simulated over a continuous 60-day horizon. Figure~\ref{comparison_traditional} presents results from day 11, comparing the proposed mechanism with the conventional double auction in terms of P2P trading price (subplot (a)) and power exchange with the utility grid (subplot (b)). Table~\ref{table_comparison} summarizes the long-term performance over the 60-day horizon, reporting the average operating cost of each microgrid and the computation time per trading interval (5 minutes) under different mechanisms.

\begin{figure*}[htbp]
  \footnotesize\rmfamily   \setlength{\abovecaptionskip}{-0.1cm}  
    \setlength{\belowcaptionskip}{-0.1cm} 
  \begin{center}  \includegraphics[width=1\columnwidth]{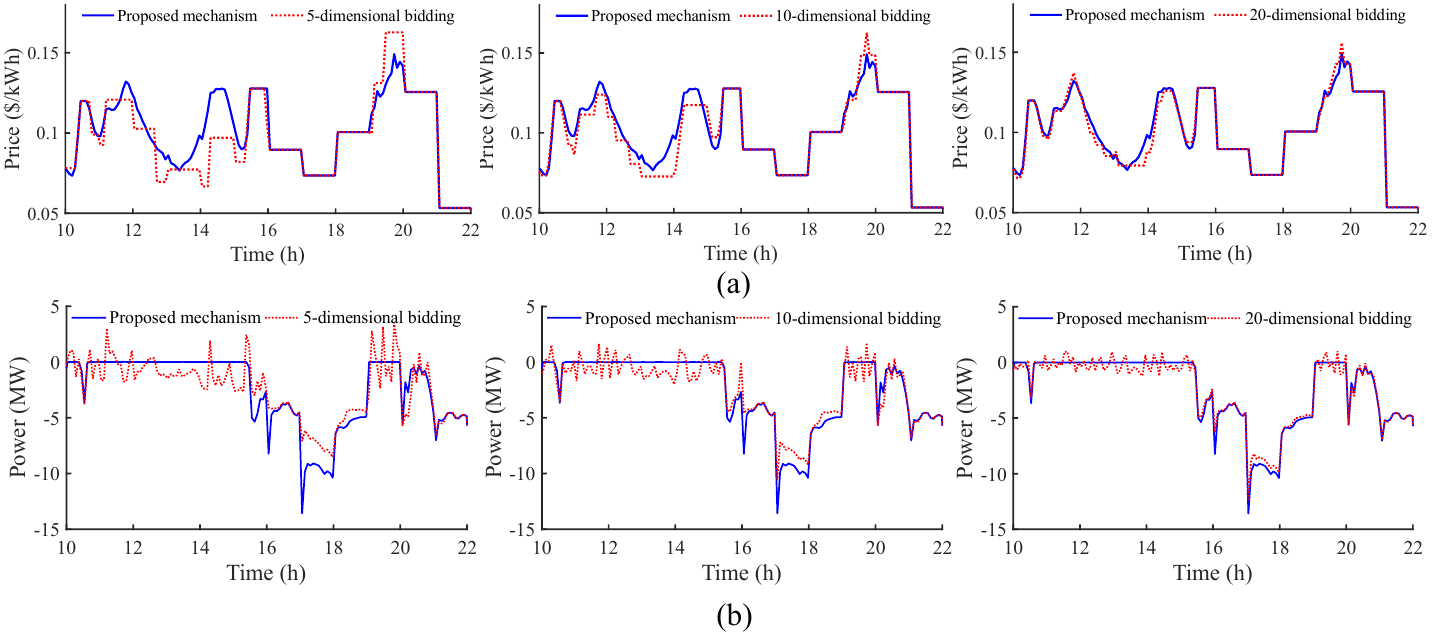}
     \caption{\rmfamily {\color{blue}Comparison between the proposed mechanism and conventional double auction under different bidding dimensions on day 11: (a) P2P trading price, (b) power exchange with the utility grid.}}
     \captionsetup{justification=centering}\label{comparison_traditional}\
  \end{center}
  \vspace{-2.5em}
\end{figure*}

\begin{table}[htbp]
\centering
\footnotesize\rmfamily
\caption{\rmfamily {\color{blue}Comparison between the proposed mechanism and conventional double auction}}
\label{table_comparison}
\begin{tabular}{ccccc}
\toprule
\textbf{{\color{blue}Metric}} & \textbf{{\color{blue}Proposed}} & \textbf{{\color{blue}5-dimensional}} & \textbf{{\color{blue}10-dimensional}} & \textbf{{\color{blue}20-dimensional}}\\
\midrule
{\color{blue}Average operating cost of each MG (\$)} & {\color{blue}14228} & {\color{blue}15131} & {\color{blue}14861} & {\color{blue}14673} \\
{\color{blue}Computation time per interval (s)} & {\color{blue}2.16} & {\color{blue}5.22} & {\color{blue}10.43} & {\color{blue}20.87} \\
\bottomrule
\end{tabular}
\end{table}

As seen in Figure~\ref{comparison_traditional} between 11:00 and 15:00, the proposed mechanism achieves exact market equilibrium through iterative updates with ASSA, yielding a clearing price that balances supply and demand. In contrast, the conventional double auction often fails to reach an exact equilibrium: although participants submit multiple price–quantity pairs, the bids are inherently discrete, making it difficult to identify a price that exactly balances supply and demand. The resulting imbalance is absorbed by the main grid, reducing the overall benefits for participating microgrids. Increasing the bidding dimensionality improves the accuracy of the clearing price and reduces operating costs, but at the expense of substantially higher computational effort for each microgrid. Table~\ref{table_comparison} confirms this trade-off: as the number of bid dimensions increases, the average operating cost decreases, but the computation time per interval rises proportionally. By comparison, the proposed mechanism simultaneously delivers the lowest operating cost and the smallest computational burden, demonstrating clear advantages over conventional double auction.}

\subsection{Performance evaluation of DDOO}

\begin{figure}[htbp]
  \footnotesize\rmfamily   \setlength{\abovecaptionskip}{-0.1cm}  
    \setlength{\belowcaptionskip}{-0.1cm} 
  \begin{center}  \includegraphics[width=0.5\columnwidth]{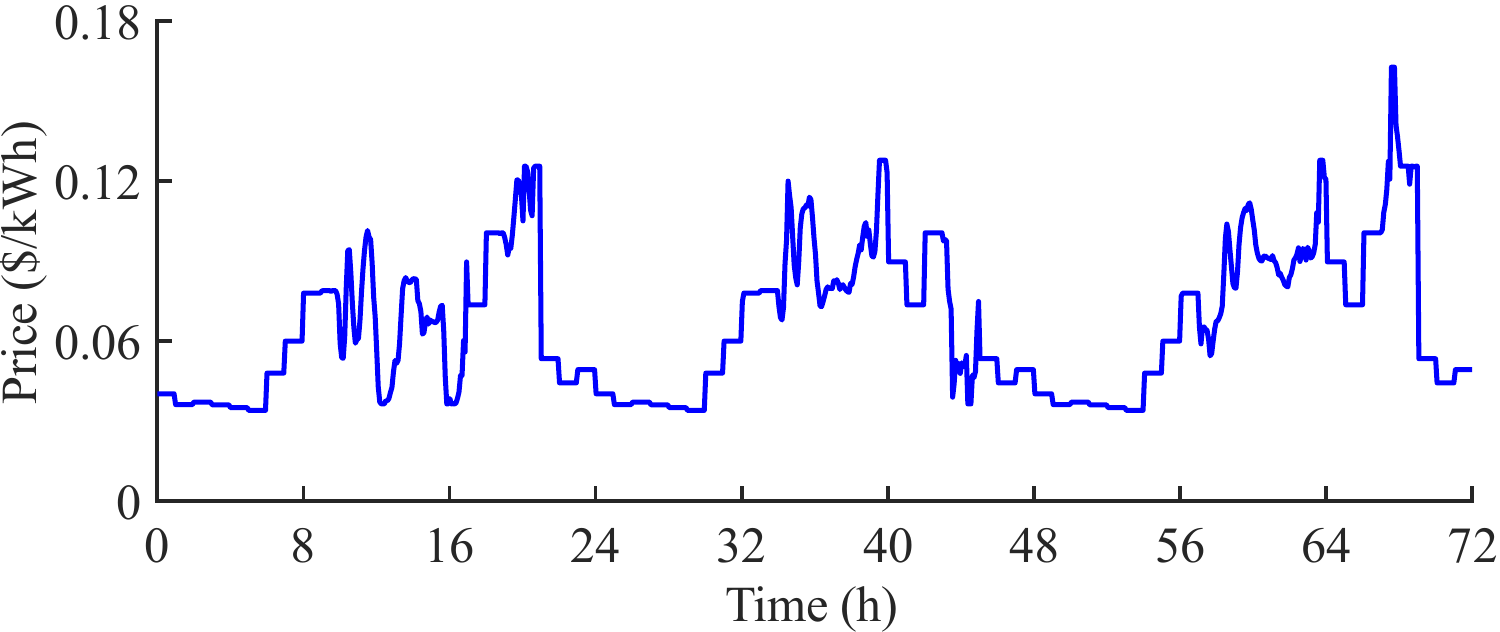}
     \caption{\rmfamily P2P trading price from day 1 to day 3. }\label{price_for_3days}
  \end{center}
  \vspace{-0.5cm}
\end{figure}
\begin{figure}[htbp]
  \footnotesize\rmfamily   \setlength{\abovecaptionskip}{-0.1cm}  
    \setlength{\belowcaptionskip}{-0.1cm} 
  \begin{center}  \includegraphics[width=0.5\columnwidth]{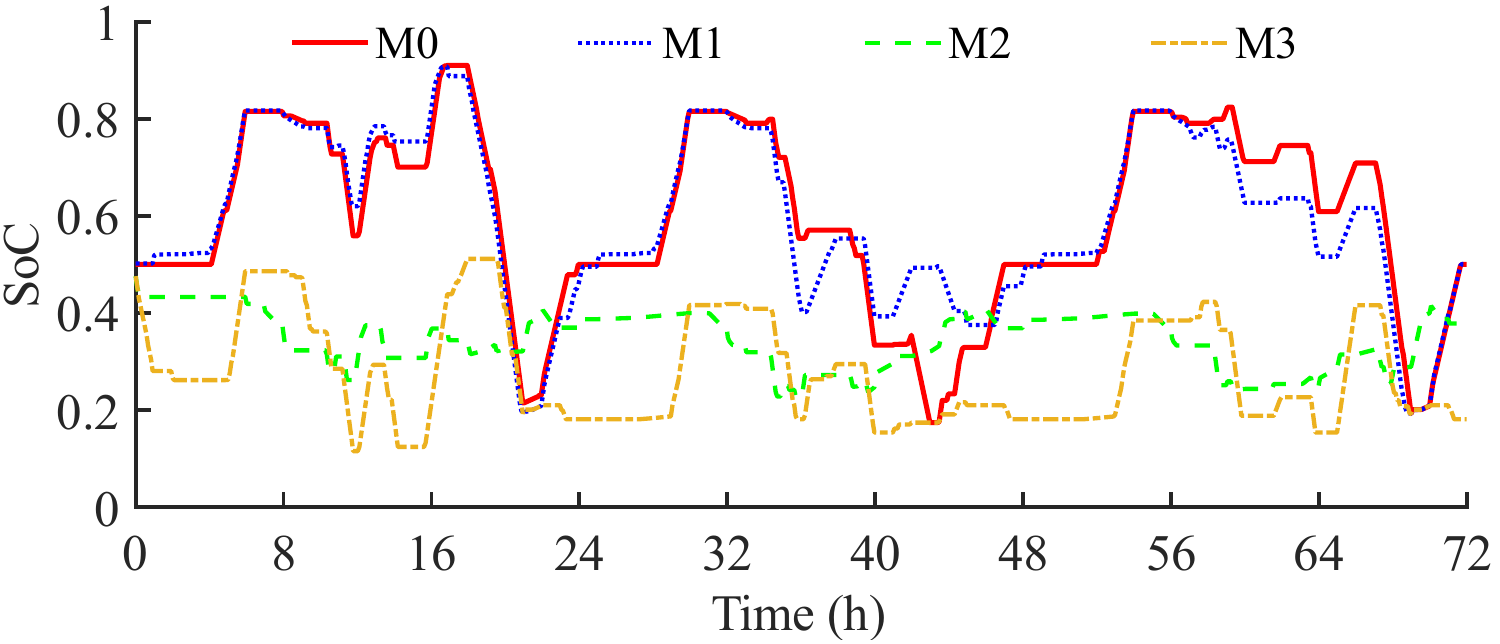}
     \caption{\rmfamily Comparison of SoC between M0-M3 for day 1-3. }\label{different_method_SoC}
  \end{center}
  \vspace{-0.5cm}
\end{figure}
\begin{figure}[htbp]
  \footnotesize\rmfamily   \setlength{\abovecaptionskip}{-0.1cm}  
    \setlength{\belowcaptionskip}{-0.1cm} 
  \begin{center}  \includegraphics[width=0.5\columnwidth]{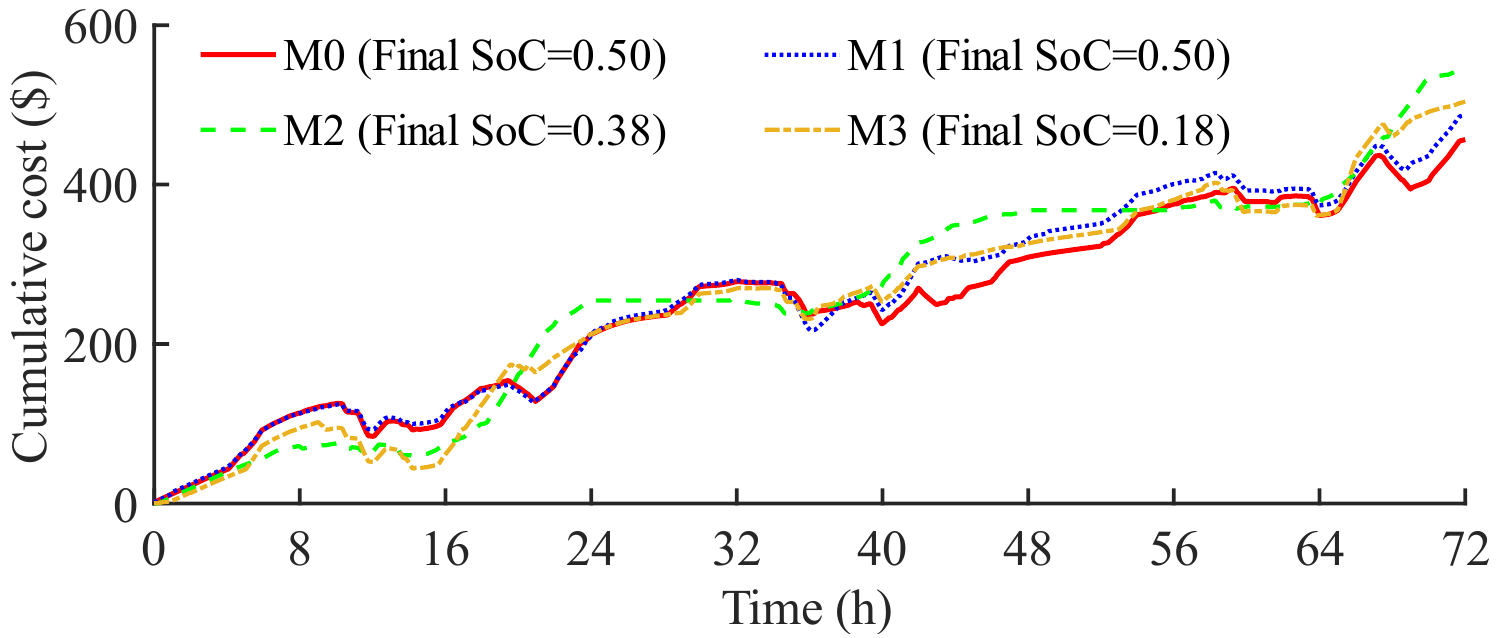}
     \caption{\rmfamily Comparison of cumulative cost between M0-M3 for day 1-3. }\label{different_method_cost}
  \end{center}
  \vspace{-0.5cm}
\end{figure}

\textcolor{blue}{In this subsection, we focus on MG7 as a representative case to investigate how different decision-making strategies affect the operation of an individual microgrid. For better comparability, we fix the market price by using the P2P trading prices obtained from the previous 60-day operation as boundary conditions. In this setting, MG7 is treated as a price-taker. This setup provides a common hindsight optimal baseline and allows us to concentrate on comparing the operation performance of MG7 under different strategies. Four approaches are considered for comparative analysis:}

\textbf{M0}: Hindsight optimization providing a theoretical lower bound. This baseline is obtained by solving the day-long optimal dispatch model with full knowledge of uncertain data, which is impractical in real-world scenarios.

\textbf{M1}: The proposed prediction-free DDOO framework.

\textbf{M2}: Lyapunov optimization~\cite{zheng2025real}, which employs the Lyapunov drift-plus-penalty technique to solve online optimization problems at each time step.

\textbf{M3}: MPC with P2P trading~\cite{shi2023distributed}. We assume that all uncertain data in the next 60 periods (four hours) can be exactly predicted and minimize the cumulative costs over the future 60 periods, but only deploy the control action for the first period. This setting is optimistic because 4-hour-ahead prediction is not accurate.

The simulation results over three consecutive days (day 1 to day 3) are illustrated in Figures~\ref{price_for_3days}–\ref{different_method_cost}. As shown in Figure~\ref{different_method_SoC}, it is evident that the SoC trajectories of M1 closely align with the ideal M0 solution, explaining its superior economic performance. In contrast, M2 consistently maintains a lower SoC due to its inherent short-sighted characteristics; specifically, its objective function only considers immediate operating costs and short-term queue stability, lacking strategic guidance for long-term energy storage management. Similarly, M3 exhibits a suboptimal behavior due to its limited prediction horizon, resulting in a relatively greedy operation that tends to maintain a low SoC level and thereby incurs higher operating costs.

Table~\ref{tab:cost_comparison} further quantifies the performance across the entire 60-day period. M1 achieves the lowest cumulative cost among the practical online methods, with only a 5.76\% optimality gap compared to the idealized M0 solution.

\begin{figure}[htbp]
  \footnotesize\rmfamily   \setlength{\abovecaptionskip}{-0.1cm}  
    \setlength{\belowcaptionskip}{-0.1cm} 
  \begin{center}  \includegraphics[width=0.5\columnwidth]{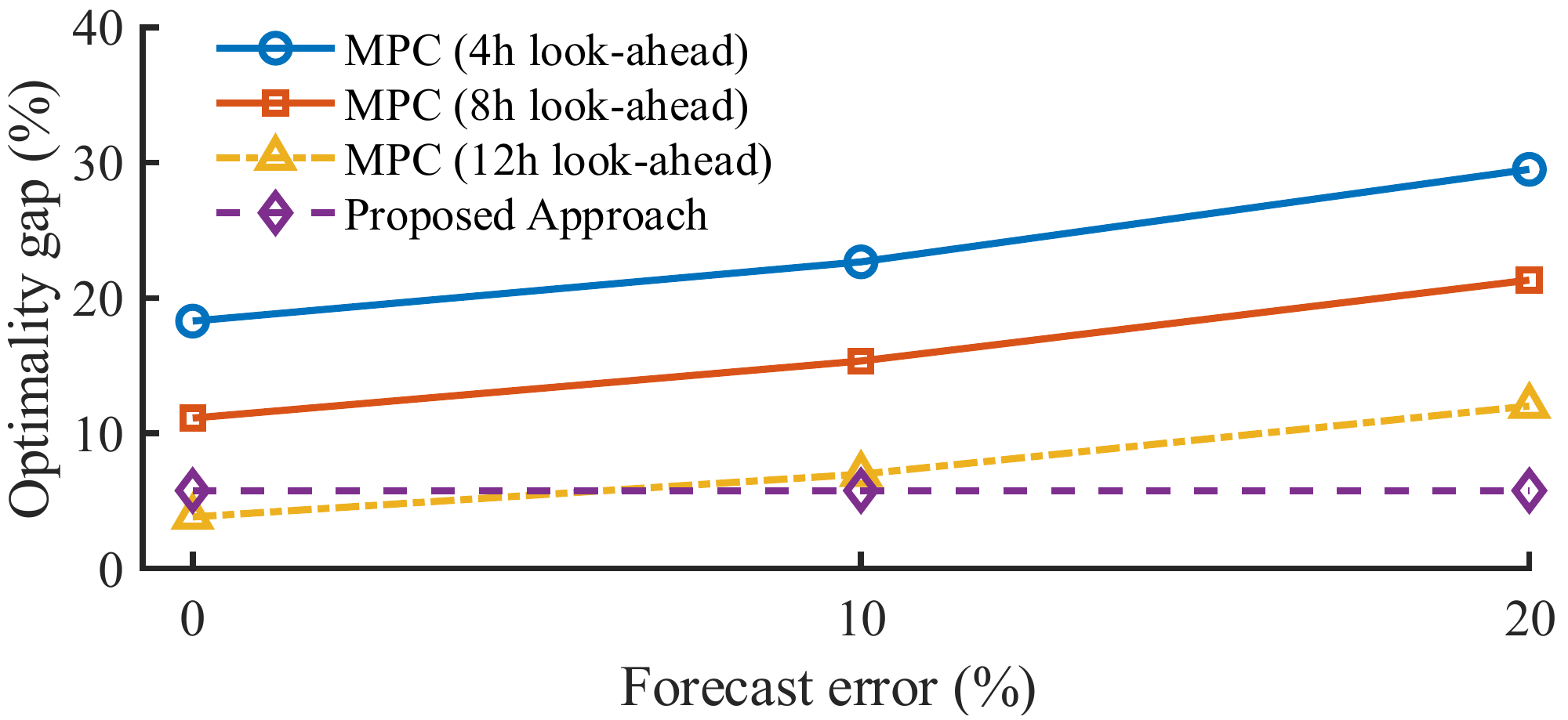}
     \caption{\rmfamily {\color{blue}Optimality gap of different methods under different forecast errors and look-ahead horizons.} }\label{different_window}
  \end{center}
  \vspace{-0.5cm}
\end{figure}

{\color{blue}
To further compare the proposed prediction-free DDOO with prediction-based strategies, we conducted an additional comparative case study using microgrid 7 over 60 consecutive days of operation. The optimality gaps of both MPC (under different look-ahead horizons and forecast error levels, measured by mean absolute percentage error) and the proposed approach are summarized in Figure~\ref{different_window}. As shown in the figure, the optimality gap of MPC grows significantly as forecast errors increase. This trend clearly demonstrates MPC’s forecast sensitivity, since its optimization directly depends on the accuracy of predicted net load and market prices. In practice, this limitation becomes even more pronounced in small-scale microgrids, where renewables are highly volatile and accurate long-horizon forecasts are particularly difficult to obtain due to the lack of reliable meteorological measurements and numerical weather predictions. In addition, Figure~\ref{different_window} also shows that the look-ahead horizon has a substantial impact on MPC performance: shorter horizons lead to noticeably larger optimality gaps. This highlights MPC’s inherent short-sightedness, as its optimization objective is restricted to minimizing costs within the finite horizon ($[t,t+\pi]$). Consequently, MPC may take actions that are locally optimal within the look-ahead window but neglect their longer-term consequences. Our case studies further confirm this limitation: even with perfect forecasts, the 8-hour MPC has an optimality gap of 11.12\%, higher than that of the proposed DDOO (5.76\%), while the 12-hour MPC achieves a lower gap of 3.81\% but relies on forecast accuracy that is unattainable in practice. By contrast, the proposed prediction-free DDOO achieves an optimality gap of only 5.76\%. This is mainly attributed to the two reference signals introduced in DDOO: the IRT, which provides long-term guidance for SoC management, and the RPB, which embeds the opportunity value of remaining storage capacity. By combining these two references, DDOO effectively avoids greedy resolutions and achieves decisions much closer to the hindsight-optimal benchmark.}

To further validate the specific contributions of the reference signals in DDOO, we evaluate two variants of the proposed method:

\textbf{M1-a}: Greedy scheduling without IRT and RPB, solely minimizing the immediate operating cost at each time step.

\textbf{M1-b}: Online optimization utilizing IRT but omitting RPB, thus not explicitly incentivizing real-time price-aware arbitrage. Specifically, this means replacing \(\hat{C}_t^\mathrm{G}\) in~\eqref{eq:modified_obj} with the original form~\eqref{GES cost}.

\begin{figure}[htbp]
  \footnotesize\rmfamily   \setlength{\abovecaptionskip}{-0.1cm}  
    \setlength{\belowcaptionskip}{-0.1cm} 
  \begin{center}  \includegraphics[width=0.5\columnwidth]{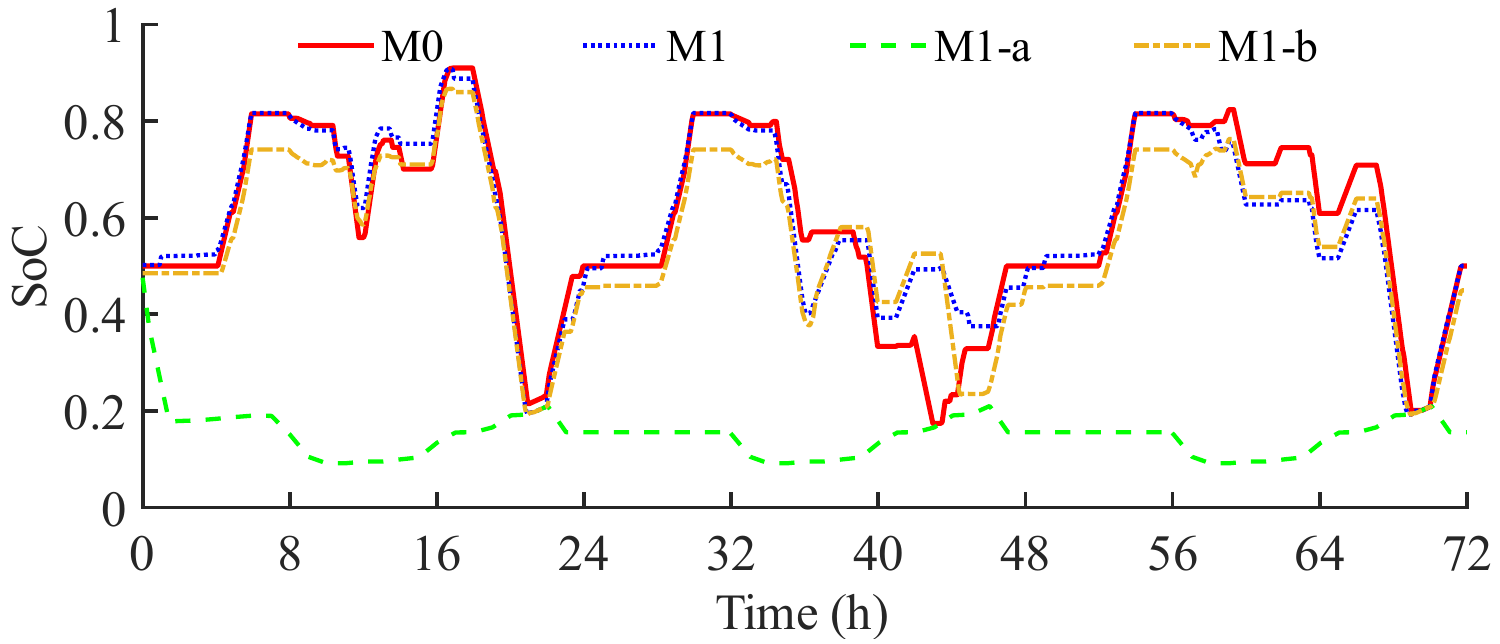}
     \caption{\rmfamily Comparison of SoC between different variants of M1 for day 1-3. }\label{different_variant_SoC}
  \end{center}
  \vspace{-0.5cm}
\end{figure}
\begin{figure}[htbp]
  \footnotesize\rmfamily   \setlength{\abovecaptionskip}{-0.1cm}  
    \setlength{\belowcaptionskip}{-0.1cm} 
  \begin{center}  \includegraphics[width=0.5\columnwidth]{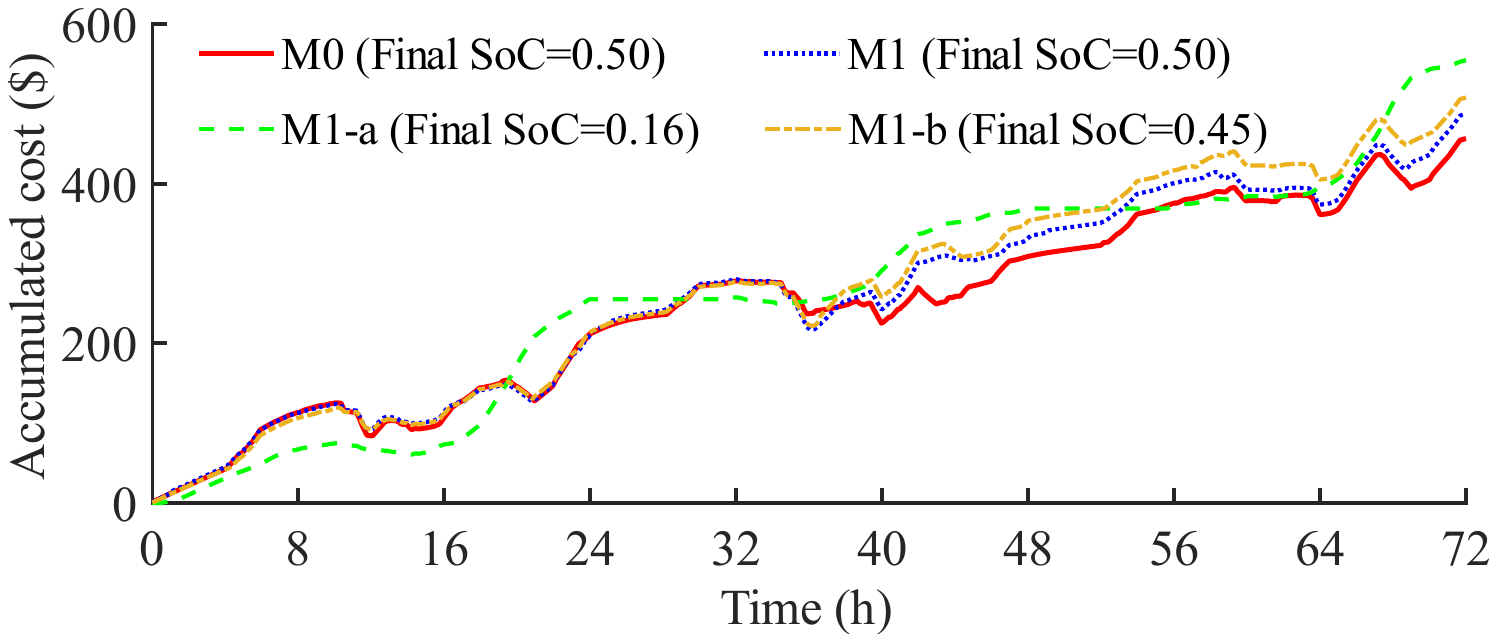}
     \caption{\rmfamily Comparison of cumulative cost between different variants of M1 for day 1-3. }\label{different_variant_cost}
  \end{center}
  \vspace{-0.5cm}
\end{figure}

\begin{table}[!b]
\centering
\footnotesize\rmfamily
\caption{\rmfamily Comparison of cumulative operating costs over 60 days}
\begin{tabular}{c p{2.8cm}<{\centering} p{2.8cm}<{\centering}}
\toprule
\textbf{Method} & \makecell{\textbf{Cumulative Cost (\$)}} & \makecell{\textbf{Optimality Gap (\%)}} \\ 
\midrule
M0  & 9187   & 0       \\
M1  & 9716   & 5.76    \\
M2  & 11135  & 21.20   \\
M3  & 10867  & 18.29   \\
M1-a & 12046 & 31.12   \\
M1-b & 10363 & 12.80   \\
\bottomrule
\end{tabular}
\label{tab:cost_comparison}
\end{table}

Figures~\ref{different_variant_SoC}–\ref{different_variant_cost} illustrate the impact of these components. The greedy M1-a rapidly depletes the GES, leading the lowest initial operating costs but the highest cumulative costs overtime, highlighting the importance of incorporating IRT. Specifically, IRT provides essential long-term strategic guidance, promoting sustainable storage operation and mitigating myopic behaviors. Meanwhile, M1-b benefits from incorporating IRT; however, its objective function still favors immediate discharging due to the absence of RPB. As a result, M1-b still exhibits higher operating costs relative to M1. This clearly demonstrates the complementary role of RPB, which provides a dynamic and instantaneous benchmark for real-time economic dispatch decisions, significantly enhancing economic performance.

{\color{blue}
For a more rigorous analysis, we further conduct a case study where MG7 participates in the market as a price-maker, that is, its operational strategy affects the market price through the auction mechanism. The results are summarized in Table~\ref{tab:cost_price_maker}. Compared with the price-taker setting, the obtained cumulative costs are very close, and the proposed DDOO framework still achieves the lowest operating cost among all practical methods.}

\begin{table}[!t]
\centering
\footnotesize\rmfamily
\caption{\rmfamily \textcolor{blue}{Comparison of cumulative operating costs over 60 days (MG7 as price-maker)}}
{\color{blue}
\begin{tabular}{c c c c c c}
\toprule
\textbf{Method} & M1 & M2 & M3 & M1-a & M1-b \\
\midrule
\textbf{Cumulative Cost (\$)} & 9840 & 11342 & 11179 & 12137 & 10458 \\
\bottomrule
\end{tabular}}
\label{tab:cost_price_maker}
\end{table}

\begin{figure}[htbp]
  \footnotesize\rmfamily   \setlength{\abovecaptionskip}{-0.1cm}  
    \setlength{\belowcaptionskip}{-0.1cm} 
  \begin{center}  \includegraphics[width=0.6\columnwidth]{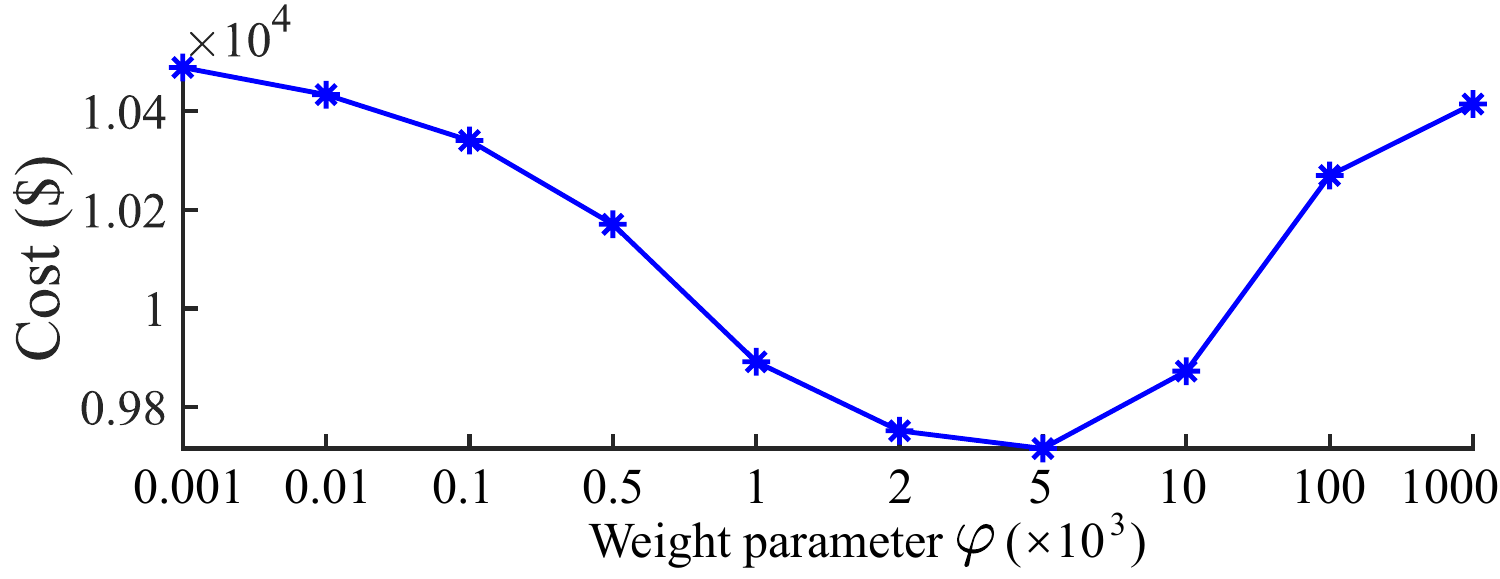}
     \caption{\rmfamily Operating cost of MG7 achieved by M1 with different $\varphi$. }\label{sensitivity_analysis}
  \end{center}
  \vspace{-3.5em}
\end{figure}

{\color{blue}
To further evaluate the robustness of the proposed DDOO framework under unexpected events, we conducted additional case studies considering sudden load spikes, renewable generation dropouts, and communication delays. 

First, we examine how a microgrid responds to sudden increases in net load, which may be caused by abrupt demand spikes or renewable generation dropouts. As shown in Figure~\ref{sensitivity_analysis}, the figure illustrates the operation of MG~12 on day~25. Two events of net load surges occur at 11:00 and 13:30 (highlighted in yellow). At 11:00, during a period of high market prices, the market price significantly exceeds the RPB, which estimates the average daily price; therefore, the microgrid promptly increases both GES discharging power and DG output to avoid expensive purchases from the market. In contrast, at 13:30, the market price is below the RPB. Although the net load spikes, the microgrid chooses to purchase power from the market instead of discharging GES. This decision proves advantageous from a hindsight perspective, as the reserved GES energy is later discharged at around 20:00 when prices peak, yielding greater economic benefit. These results demonstrate that DDOO can effectively respond to real-time changes in net power while making advantageous decisions based on the prevailing market price.}

\begin{figure}[htbp]
  \footnotesize\rmfamily   \setlength{\abovecaptionskip}{-0.1cm}  
    \setlength{\belowcaptionskip}{-0.1cm} 
  \begin{center}  \includegraphics[width=0.7\columnwidth]{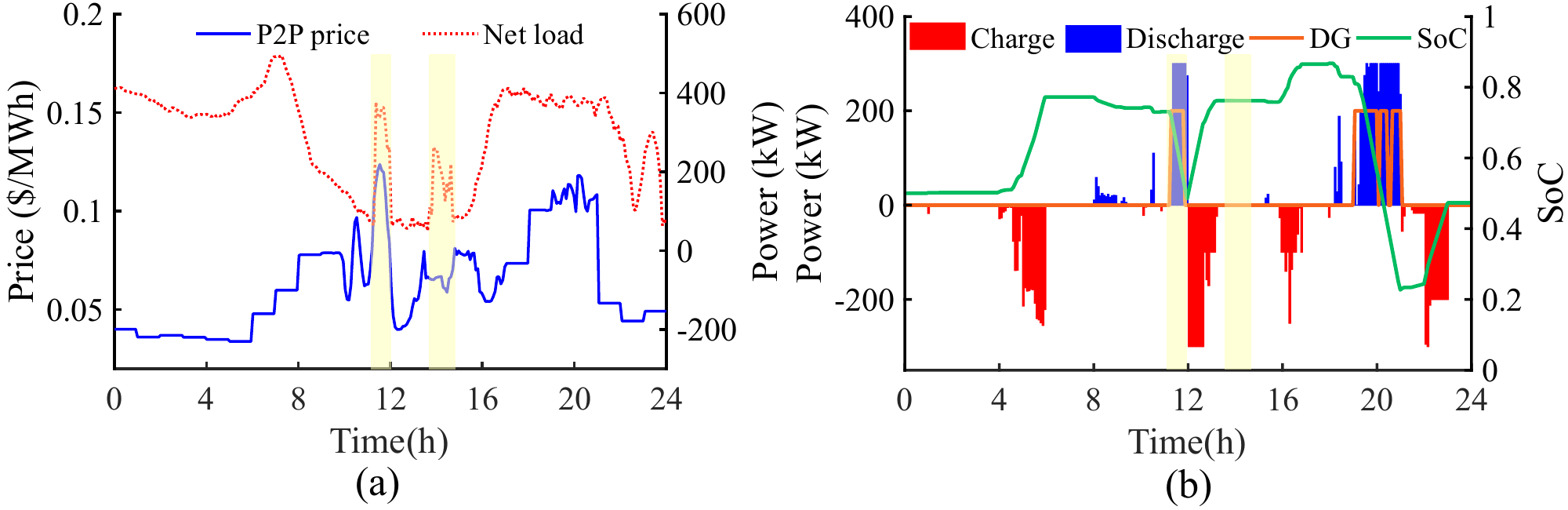}
     \caption{\rmfamily {\color{blue}Operational results of MG12 on day~25: (a) Net load and market price; (b) Energy management strategies.} }\label{robust_cas-refs-1check}
  \end{center}
  \vspace{-3.5em}
\end{figure}

{\color{blue}Second, since microgrid decisions rely on real-time measurements, communication delays may affect DDOO performance. In practice, however, communication latency in microgrids is typically in the order of tens to hundreds of milliseconds, and even in wireless networks usually below one second. During such short communication delays, changes in renewable generation and load power are typically negligible. Given the minute-level timescale of DDOO operation, such sub-second delays are negligible relative to the scheduling timescale. To investigate more extreme cases, we further test scenarios where input signals are delayed by 1–2 dispatch intervals (i.e., 5–10 minutes). Taking MG~7 as an example, Table~\ref{table_delay} reports the cumulative operating cost over a 60-day horizon under different delay conditions.}

\begin{table}[htbp]
\centering
\footnotesize\rmfamily
\caption{\rmfamily {\color{blue}Cumulative operating cost of MG~7 over 60 days under different communication delays}}
\label{table_delay}
\begin{tabular}{ccc}
\toprule
\textbf{{\color{blue}Scenario}} & \textbf{{\color{blue}Cumulative Cost (\$)}} & \textbf{{\color{blue}Optimality Gap (\%)}} \\
\midrule
{\color{blue}No delay} & {\color{blue}9716} & {\color{blue}5.76}\\
{\color{blue}5-min delay} & {\color{blue}10271} & {\color{blue}11.80}\\
{\color{blue}10-min delay} & {\color{blue}10947} & {\color{blue}19.16}\\
\bottomrule
\end{tabular}
\end{table}

{\color{blue}
The results show that while communication delays increase the operating cost to some extent, the overall degradation remains moderate and does not lead to severe performance loss. Moreover, such long delays are rare in practical microgrid environments. This indicates that the proposed DDOO framework possesses good robustness against communication latency. In practice, if very high delays or packet losses occur due to device failures or exceptional network conditions, simple prediction-based methods, such as trend extrapolation or moving averages, can be employed to compensate for missing data and further enhance the resilience of the dispatch strategy.}

Finally, a sensitivity analysis on the weighting parameter $\varphi$ in~\eqref{eq:modified_obj} is provided in Figure~\ref{sensitivity_analysis}. This parameter determines how closely the actual SoC is constrained to track the IRT. As $\varphi$ increases, the operating cost initially decreases and subsequently increases, with a moderate value ($\varphi=5000$) yielding the lowest cost. This phenomenon occurs because, although IRT provides strategic guidance, it inevitably differs from the optimal SoC. Hence, RPB-based real-time adjustments provide necessary operational flexibility to minimize immediate costs and capitalize on peak-valley arbitrage opportunities.

\section{Conclusion}~\label{conclusion}

This paper proposed an improved double auction mechanism combined with an ASSA and a prediction-free DDOO framework to address {\color{blue}computational complexity and myopic decision-making} in real-time P2P energy trading. Comprehensive case studies verify that:

(1) The proposed market mechanism significantly enhances local RES consumption and economic efficiency. Compared with the case without P2P trading, the self-sufficiency period increased from 0.01\% to 29.86\%, the reverse power flow period reduced from 24.51\% to 3.96\%, and the average operating cost of each MG decreased by 19.20\%.

(2) The ASSA ensures rapid convergence to market equilibrium. On average, convergence is achieved within 2.07 iterations per 5-min interval, corresponding to a computational speedups of approximately 9.68 and 4.84 times compared to traditional 20-dimensional and 10-dimensional bidding approaches, respectively.

(3) The proposed DDOO framework significantly alleviates the myopic limitations of conventional online optimization methods. Compared with MPC-based and Lyapunov optimization methods, DDOO reduced operating cost by approximately 10.59\% and 12.74\%, respectively, and achieved an optimality gap of just 5.76\% relative to the ideal hindsight solution.

Future work will extend the proposed method to integrate a broader range of heterogeneous resources and energy carriers, such as long-duration ES and thermal energy systems, and will also incorporate carbon market mechanisms.
\newpage

\begin{center}
     \textbf{Appendix}
\end{center}

\begin{appendices}
\section*{Appendix A. Proof of Proposition~\ref{Proposition1}}\label{appendix_A}

At any given trading interval $t$ and for any MG $r$, the optimal response function is defined in~\eqref{eq:optimal_response_definition}, where the internal cost function \( f_{t,r}(P_{t,r}^{\mathrm{EX}},\lambda_t^{\mathrm{P2P}}) \) is explicitly given in~\eqref{eq:original_objective}.

The SoC dynamics of GES are described by:
\begin{equation}
SoC_t = SoC_{t-1} + \frac{\eta^{\mathrm{c}} P_{t}^\mathrm{G,c}\Delta t - P_{t}^\mathrm{G,d}\Delta t/\eta^{\mathrm{d}}}{E}+\pi_t,\label{eq:SoC_dynamics}
\end{equation}
and the net export power is constrained by power balance:
\begin{equation}
P_{t,r}^{\mathrm{EX}} = P_{t}^{\mathrm{L}}-P_{t}^{\mathrm{R}}+P_{t}^{\mathrm{G,c}}-P_{t}^{\mathrm{G,d}}-P_{t}^{\mathrm{DG}}.\label{eq:power_balance_appendix}
\end{equation}

Substituting \eqref{eq:SoC_dynamics} and \eqref{eq:power_balance_appendix} into \eqref{eq:original_objective}, and denoting $P_{t,r}^{\mathrm{EX}}$ explicitly as the decision variable, the MG cost function can be rewritten in a simplified quadratic form:
\begin{equation}
f_{t,r}(P_{t,r}^{\mathrm{EX}},\lambda_t^{\mathrm{P2P}})=c_t^{\mathrm{B}}(P_{t,r}^{\mathrm{EX}})^2 + c_t^{\mathrm{A}}P_{t,r}^{\mathrm{EX}} + c_t^{0},\label{eq:simplified_objective}
\end{equation}
where:
\begin{equation}
c_t^{\mathrm{B}}=\frac{\varphi(\Delta t)^2}{E^2(\eta^{\mathrm{d}})^2}>0,
\end{equation}

Clearly, the function in~\eqref{eq:simplified_objective} is strictly convex with respect to $P_{t,r}^{\mathrm{EX}}$ due to $c_t^{\mathrm{B}}>0$. Hence, the global unconstrained optimal solution can be obtained by setting the first-order derivative to zero:
\begin{equation}
\frac{\partial f_{t,r}}{\partial P_{t,r}^{\mathrm{EX}}}=2c_t^{\mathrm{B}}P_{t,r}^{\mathrm{EX}} + c_t^{\mathrm{A}}=0,
\end{equation}
resulting in:
\begin{equation}
P_{t,r}^{\mathrm{EX,*}}(\lambda_t^{\mathrm{P2P}})=-\frac{c_t^{\mathrm{A}}}{2c_t^{\mathrm{B}}}.\label{eq:unconstrained_optimal}
\end{equation}

The derivative of the optimal response function with respect to $\lambda_t^{\mathrm{P2P}}$ is thus:
\begin{equation}
\frac{\mathrm{d}P_{t,r}^{\mathrm{EX,*}}}{\mathrm{d}\lambda_t^{\mathrm{P2P}}}=-\frac{1}{2c_t^{\mathrm{B}}}<0.\label{eq:derivative_unconstrained}
\end{equation}

Equation~\eqref{eq:derivative_unconstrained} demonstrates that the unconstrained optimal response function is strictly monotonically decreasing and continuous.

In practical market operations, MGs' export power is typically bounded by physical and operational constraints, which correspond precisely to constraints~\eqref{DG_output}--\eqref{GES_leq},~\eqref{GES_SOC}, and~\eqref{P_balance}--\eqref{reformation} in the dispatch model~\eqref{eq:modified_obj}. These constraints collectively imply that the feasible trading power at any given time interval lies within a bounded range, compactly expressed as:
\begin{equation}
P_{t,r}^{\mathrm{EX,min}} \leq P_{t,r}^{\mathrm{EX}} \leq P_{t,r}^{\mathrm{EX,max}}.\label{eq:export_constraints}
\end{equation}

Considering these constraints, the optimal response becomes the projection of the unconstrained optimal solution onto the feasible set defined by~\eqref{eq:export_constraints}:
\begin{equation}
P_{t,r}^{\mathrm{EX,*}}(\lambda_t^{\mathrm{P2P}})=\left[-\frac{c_t^{\mathrm{A}}}{2c_t^{\mathrm{B}}}\right]_{P_{t,r}^{\mathrm{EX,min}}}^{P_{t,r}^{\mathrm{EX,max}}}.
\end{equation}

Since the projection operator is non-decreasing and the unconstrained solution is strictly decreasing, the constrained optimal response function retains the continuity and monotonic non-increasing properties. Although there might exist horizontal segments due to boundary constraints, the function as a whole remains continuous and monotonically decreasing.

Therefore, the optimal response function $P_{t,r}^{\mathrm{EX,*}}(\lambda_t^{\mathrm{P2P}})$ is continuous and monotonically decreasing in $\lambda_t^{\mathrm{P2P}}$, completing the proof of Proposition~\ref{Proposition1}.

\section*{Appendix B. Proof of Proposition~\ref{Proposition2}}\label{appendix_B}

Define the aggregate supply-demand function as:
\begin{equation}\label{eq:aggregate_supply_demand}
F_t(\lambda_t^{\mathrm{P2P}})=\sum_{r=1}^{N^r} P_{t,r}^{\mathrm{EX,*}}(\lambda_t^{\mathrm{P2P}}).
\end{equation}

Due to the continuity and monotonic non-increasing properties of each MG's optimal response \(P_{t,r}^{\mathrm{EX,*}}\), it immediately follows that \(F_t(\lambda_t^{\mathrm{P2P}})\) is continuous and monotonically non-increasing over the interval \([\lambda_t^{\mathrm{FiT}},\lambda_t^{\mathrm{ToU}}]\).

Let the range of \(F_t\) within the domain \([\lambda_t^{\mathrm{FiT}},\lambda_t^{\mathrm{ToU}}]\) be \([F_{t,\min}, F_{t,\max}]\), where:
\begin{subequations}
\begin{align}
F_{t,\min}&=F_t(\lambda_t^{\mathrm{ToU}}),\\[3pt]
F_{t,\max}&=F_t(\lambda_t^{\mathrm{FiT}}).
\end{align}
\end{subequations}

Three distinct cases are possible:

\textbf{Case 1:} If \(F_{t,\min}\leq 0 \leq F_{t,\max}\), then due to continuity and monotonicity, the Intermediate Value Theorem guarantees the existence of a unique \(\lambda_t^{\mathrm{P2P}*}\in[\lambda_t^{\mathrm{FiT}},\lambda_t^{\mathrm{ToU}}]\) satisfying:
\begin{equation}
F_t(\lambda_t^{\mathrm{P2P}*})=0.
\end{equation}
Here, the equilibrium conditions~\eqref{eq:price_bound}--\eqref{eq:grid_trade} hold, with \(P_t^{\mathrm{grid}}=0\), and uniqueness is evident from strict monotonicity.

\textbf{Case 2:} If \(F_{t,\max}<0\), even at the lowest feasible price \(\lambda_t^{\mathrm{FiT}}\), the P2P market experiences excess supply. Therefore, energy is exported to the utility grid, and the grid exchange power is:
\begin{equation}
P_t^{\mathrm{grid}}=-F_t(\lambda_t^{\mathrm{FiT}})>0.
\end{equation}
In this scenario, conditions~\eqref{eq:price_bound}--\eqref{eq:grid_trade} still hold at the unique equilibrium price:
\begin{equation}
\lambda_t^{\mathrm{P2P}*}=\lambda_t^{\mathrm{FiT}}.
\end{equation}

\textbf{Case 3:} If \(F_{t,\min}>0\), even at the highest feasible price \(\lambda_t^{\mathrm{ToU}}\), the P2P market experiences excess demand. Thus, energy must be imported from the utility grid, resulting in:
\begin{equation}
P_t^{\mathrm{grid}}=-F_t(\lambda_t^{\mathrm{ToU}})<0.
\end{equation}
All equilibrium conditions~\eqref{eq:price_bound}--\eqref{eq:grid_trade} remain satisfied at the unique equilibrium price:
\begin{equation}
\lambda_t^{\mathrm{P2P}*}=\lambda_t^{\mathrm{ToU}}.
\end{equation}

Therefore, for all three exhaustive scenarios, the equilibrium satisfying conditions~\eqref{eq:price_bound}--\eqref{eq:grid_trade} exists and is unique. This completes the proof.

\section*{Appendix C. Proof of Proposition~\ref{Proposition3}}\label{appendix_C}

We analyze the convergence of the proposed ASSA algorithm in three distinct scenarios.

\textbf{Scenario 1 (Excess Demand):} Suppose that within the feasible price interval $[\lambda_t^{\mathrm{FiT}}, \lambda_t^{\mathrm{ToU}}]$, the aggregate supply-demand function $F_t(\lambda_t^{\mathrm{P2P}})$ satisfies:
\begin{equation}
F_t(\lambda_t^{\mathrm{P2P}}) > 0,\quad \forall \lambda_t^{\mathrm{P2P}} \in [\lambda_t^{\mathrm{FiT}}, \lambda_t^{\mathrm{ToU}}].
\end{equation}

In this scenario, the P2P market always exhibits excess demand. According to Algorithm~\ref{algorithm_ASSA}, the price updating rule is:
\begin{equation}
\lambda_{k+1,t}^{\mathrm{P2P}} = \lambda_{k,t}^{\mathrm{P2P}} + \sigma F_t(\lambda_{k,t}^{\mathrm{P2P}}).
\end{equation}

Since $F_t(\lambda_{k,t}^{\mathrm{P2P}}) > 0$, each iteration strictly increases the price. Additionally, there will be no sign changes in $F_t(\lambda_t^{\mathrm{P2P}})$, thus the step size $\sigma$ remains constant during iterations. Given the continuity and strict positivity of $F_t(\lambda_t^{\mathrm{P2P}})$, the minimum increment per iteration is bounded by:
\begin{equation}
\lambda_{k+1,t}^{\mathrm{P2P}} - \lambda_{k,t}^{\mathrm{P2P}} = \sigma F_t(\lambda_{k,t}^{\mathrm{P2P}}) \geq \sigma F_t(\lambda_t^{\mathrm{ToU}}) > 0.
\end{equation}

Consequently, the market-clearing price increases monotonically until it reaches the upper boundary price $\lambda_t^{\mathrm{ToU}}$. Thus, the maximum required number of iterations is explicitly bounded by:
\begin{equation}
K_{\max} = \left\lceil\frac{\lambda_t^{\mathrm{ToU}} - \lambda_{0,t}^{\mathrm{P2P}}}{\sigma F_t(\lambda_t^{\mathrm{ToU}})}\right\rceil.
\end{equation}

Therefore, ASSA converges to the boundary equilibrium price $\lambda_t^{\mathrm{ToU}}$ within a finite number of iterations not exceeding $K_{\max}$.

\textbf{Scenario 2 (Excess Supply):} Suppose that within the feasible price interval $[\lambda_t^{\mathrm{FiT}}, \lambda_t^{\mathrm{ToU}}]$, the aggregate supply-demand function $F_t(\lambda_t^{\mathrm{P2P}})$ satisfies:
\begin{equation}
F_t(\lambda_t^{\mathrm{P2P}}) < 0,\quad \forall \lambda_t^{\mathrm{P2P}} \in [\lambda_t^{\mathrm{FiT}}, \lambda_t^{\mathrm{ToU}}].
\end{equation}

In this scenario, the market consistently has excess supply. Similarly, each iteration strictly decreases the price. Again, no step-size halving occurs due to the absence of sign changes. Given the continuity and strict negativity of $F_t(\lambda_t^{\mathrm{P2P}})$, the minimum decrement per iteration is bounded by:
\begin{equation}
\lambda_{k,t}^{\mathrm{P2P}} - \lambda_{k+1,t}^{\mathrm{P2P}} = -\sigma F_t(\lambda_{k,t}^{\mathrm{P2P}}) \geq -\sigma F_t(\lambda_t^{\mathrm{FiT}}) > 0.
\end{equation}

Thus, the market-clearing price monotonically decreases until it reaches the lower boundary price $\lambda_t^{\mathrm{FiT}}$. Hence, the maximum number of iterations required is explicitly bounded by:
\begin{equation}
K_{\max} = \left\lceil\frac{\lambda_{0,t}^{\mathrm{P2P}} - \lambda_t^{\mathrm{FiT}}}{-\sigma F_t(\lambda_t^{\mathrm{FiT}})}\right\rceil.
\end{equation}

\textbf{Scenario 3 (Existence of Equilibrium):} Suppose that within the feasible price interval \([\lambda_t^{\mathrm{FiT}}, \lambda_t^{\mathrm{ToU}}]\), the aggregate supply-demand function \(F_t(\lambda_t^{\mathrm{P2P}})\) is continuous, monotonically decreasing, and there exists a unique equilibrium price \(\lambda_t^{\mathrm{P2P,*}}\in(\lambda_t^{\mathrm{FiT}}, \lambda_t^{\mathrm{ToU}})\) satisfying:
\[
F_t(\lambda_t^{\mathrm{P2P,*}})=0.
\]

We now prove that the ASSA algorithm converges to the equilibrium price \(\lambda_t^{\mathrm{P2P,*}}\) within a finite number of iterations, employing fixed-point iteration theory as the theoretical foundation.

\textit{Step 1 (ASSA as Fixed-point Iteration):}

The ASSA iteration can be expressed as a fixed-point iteration of the form:
\begin{equation}
\lambda_{k+1,t}^{\mathrm{P2P}} = g_t(\lambda_{k,t}^{\mathrm{P2P}}),\quad\text{where}\ g_t(\lambda)=\lambda+\sigma F_t(\lambda).
\end{equation}

Clearly, the equilibrium price \(\lambda_t^{\mathrm{P2P,*}}\) is the fixed point, as:
\begin{equation}
g_t(\lambda_t^{\mathrm{P2P,*}})=\lambda_t^{\mathrm{P2P,*}}+\sigma F_t(\lambda_t^{\mathrm{P2P,*}})=\lambda_t^{\mathrm{P2P,*}}.
\end{equation}

According to the classical fixed-point iteration convergence theorem~\cite{berinde2007iterative,HoelFixedPoint}, if the iteration function \( g_t(\lambda) \) is continuously differentiable around the fixed point \( \lambda_t^{\mathrm{P2P,*}} \), and satisfies:
\begin{equation}
|g_t'(\lambda_t^{\mathrm{P2P,*}})|<1,
\end{equation}
then there exists a local neighborhood \(U(\lambda_t^{\mathrm{P2P,*}},\epsilon)=\{\lambda:\,|\lambda-\lambda_t^{\mathrm{P2P,*}}|<\epsilon\}\), such that any iteration initiated within this neighborhood converges to the fixed point \(\lambda_t^{\mathrm{P2P,*}}\).

\textit{Step 2 (Local Convergence Condition):}

The derivative of \(g_t(\lambda)\) at the equilibrium \(\lambda_t^{\mathrm{P2P,*}}\) is:
\begin{equation}
g_t'(\lambda_t^{\mathrm{P2P,*}})=1+\sigma F_t'(\lambda_t^{\mathrm{P2P,*}}).
\vspace{-0.1cm}
\end{equation}

Given the strictly monotonically decreasing property of \(F_t(\lambda)\), we have \(F_t'(\lambda_t^{\mathrm{P2P,*}})<0\). Thus, to ensure:
\begin{equation}
|g_t'(\lambda_t^{\mathrm{P2P,*}})|=|1+\sigma F_t'(\lambda_t^{\mathrm{P2P,*}})|<1,
\end{equation}
it suffices to choose a sufficiently small step size satisfying:
\vspace{-0.1cm}
\begin{equation}\label{step_condition}
0<\sigma<\frac{2}{|F_t'(\lambda_t^{\mathrm{P2P,*}})|}.
\vspace{-0.2cm}
\end{equation}

Hence, under this step-size condition, there exists a local neighborhood \(U(\lambda_t^{\mathrm{P2P,*}},\epsilon)\) around the equilibrium within which the iteration is guaranteed to converge.

\textit{Step 3 (Global Convergence via Adaptive Step-Size):}

We now prove that starting from any initial point \(\lambda_{0,t}^{\mathrm{P2P}}\in[\lambda_t^{\mathrm{FiT}},\lambda_t^{\mathrm{ToU}}]\) with positive initial step size \(\sigma_0>0\), the ASSA iteration sequence must enter the local convergence region \(U(\lambda_t^{\mathrm{P2P,*}},\epsilon)\) within a finite number of iterations.

Given the strictly decreasing property of \(F_t(\lambda)\), we analyze two possible scenarios:

- \textit{Case A (No oscillation occurs):} In this case, the sequence monotonically approaches the equilibrium \(\lambda_t^{\mathrm{P2P,*}}\). Specifically, if \(\lambda_{k,t}^{\mathrm{P2P}}>\lambda_t^{\mathrm{P2P,*}}\), then \(F_t(\lambda_{k,t}^{\mathrm{P2P}})<0\) implies \(\lambda_{k+1,t}^{\mathrm{P2P}}<\lambda_{k,t}^{\mathrm{P2P}}\), causing a monotonic decreasing sequence approaching from the right. Similarly, if \(\lambda_{k,t}^{\mathrm{P2P}}<\lambda_t^{\mathrm{P2P,*}}\), the sequence monotonically increases toward the equilibrium from the left side. Due to monotonicity and boundedness, the sequence will necessarily enter the local neighborhood \(U(\lambda_t^{\mathrm{P2P,*}},\epsilon)\) after finite iterations.

- \textit{Case B (Oscillation occurs):} Oscillation occurs when two consecutive iterations lie on opposite sides of the equilibrium, causing:
\begin{equation}
F_t(\lambda_{k,t}^{\mathrm{P2P}})\cdot F_t(\lambda_{k-1,t}^{\mathrm{P2P}})<0,
\end{equation}
triggering the ASSA step-size halving rule: $\sigma \leftarrow \sigma / 2$;

Each oscillation thus halves the step size, and since there can only be a finite number of such oscillations before the step size becomes sufficiently small, we must reach a step size that satisfies the local convergence condition~\eqref{step_condition}.

Once this occurs, subsequent iterations will no longer cross the equilibrium excessively and will monotonically approach the equilibrium from one side. As in Case A, the iteration must then enter the local convergence region after finitely many further steps.

Therefore, ASSA rigorously ensures convergence within finite iterations, i.e., there exists a finite integer \(k^*\) such that:
\begin{equation}
|F_t(\lambda_{k^*,t}^{\mathrm{P2P}})|\leq\delta,
\vspace{-0.1cm}
\end{equation}
where \(\delta>0\) is the predefined convergence tolerance. 

This completes the rigorous convergence proof.
\vspace{-0.3cm}

\end{appendices}

\printcredits
\vspace{0.3cm}

\noindent \textbf{Declaration of competing interest}

Given his role as the Managing Editor of \textit{Applied Energy}, Yue Zhou had no involvement in the peer review of this article and had no access to information regarding its peer review. Full responsibility for the editorial process for this article was delegated to another journal editor. If there are other authors, they declare that they have no known competing financial interests or personal relationships that could have appeared to influence the work reported in this paper.
\vspace{0.3cm}

\noindent \textbf{Data availability}

The original data can be downloaded from~\cite{huangdata}.
\vspace{0.2cm}

\noindent \textbf{Acknowledgements}

This work was supported by the Innovation Support Program (Soft Science Research) of Jiangsu Province of China under Grant BE2023093-1 and China Postdoctoral Science Foundation special funded project (No. 2023TQ0169). We also thank Associate Professor Feng Liu from the Department of Electrical Engineering, Tsinghua University, for his course on “Engineering Game Theory,” which provided valuable insights for this work.
\vspace{-0.2cm}

\bibliographystyle{unsrt}

\bibliography{cas-refs-1}



\end{document}